\documentclass[journal=jacsat,manuscript=article]{achemso}

\usepackage[version=3]{mhchem} 
\usepackage{xcolor}
\usepackage{amsmath}
\usepackage{amssymb}
\usepackage{relsize}
\usepackage{academicons} 
\usepackage{hyperref}
\hypersetup{
    colorlinks=true,
    linkcolor=black,
    citecolor=black,
    urlcolor=black
}

\setkeys{acs}{doi=true}

\newcommand{\angstrom}{\textup{\AA}}
\newcommand{\orcid}[1]{\href{https://orcid.org/#1}{\textcolor{green}{\aiOrcid}}}

\newcommand{\beginsupplement}{%
    \setcounter{table}{0}
    \renewcommand{\thetable}{S\arabic{table}}%
    \setcounter{page}{1}
    \renewcommand{\thepage}{S\arabic{page}}
    \setcounter{figure}{0}
    \renewcommand{\thefigure}{S\arabic{figure}}%
    \setcounter{equation}{0}
    \renewcommand{\theequation}{S\arabic{equation}}%
    \setcounter{section}{0}
    \renewcommand{\thesection}{S\arabic{section}}%
    \setcounter{subsection}{0}%
    \renewcommand{\thesubsection}{S\arabic{section}.\arabic{subsection}}%
    \newcounter{SItab}
    \renewcommand{\theSItab}{S\arabic{SItab}}%
    \newcounter{SIfig}
    \renewcommand{\theSIfig}{S\arabic{SIfig}}%
}

\author{Marta Monti}
\affiliation[ICTP]{Condensed Matter and Statistical Physics, The Abdus Salam International Centre for Theoretical Physics (ICTP), Trieste 34151, Italy}
\email{mmonti@ictp.it}
\author{Luca Cimmino}
\affiliation[UniNa1]{IRCCS SYNLAB SDN, 80146 Naples, Italy}
\author{Gonzalo Díaz Mirón}
\affiliation[ICTP]{Condensed Matter and Statistical Physics, The Abdus Salam International Centre for Theoretical Physics (ICTP), Trieste 34151, Italy}
\author{Carlo Diaferia}
\affiliation[UniNa2]
{Department of Pharmacy, Research Centre on Bioactive Peptides (CIRPeB), University of Naples “Federico II”, Naples, 80134 Italy}
\author{Debarshi Banerjee}
\affiliation[ICTP]{Condensed Matter and Statistical Physics, The Abdus Salam International Centre for Theoretical Physics (ICTP), Trieste 34151, Italy}
\alsoaffiliation[SISSA]{Molecular and Statistical Biophysics, Scuola Internazionale Superiore di Studi Avanzati (SISSA), Via Bonomea 265, 34136 Trieste, Italy}
\author{Martina Stella}
\affiliation[ICTP]{Condensed Matter and Statistical Physics, The Abdus Salam International Centre for Theoretical Physics (ICTP), Trieste 34151, Italy}
\alsoaffiliation[Algo]{Algorithmiq Ltd, Kanavakatu, 3C, Fl-00160, Helsinki, 00160, Finland}
\author{Luigi Vitagliano}
\affiliation[CNR]
{Institute of Biostructures and Bioimaging, CNR, 80131 Naples, Italy}
\author{Antonella Accardo}
\affiliation[UniNa2]
{Department of Pharmacy, Research Centre on Bioactive Peptides (CIRPeB), University of Naples “Federico II”, Naples, 80134 Italy}
\email{antonella.accardo@unina.it}
\author{Ali Hassanali}
\affiliation[ICTP]{Condensed Matter and Statistical Physics, The Abdus Salam International Centre for Theoretical Physics (ICTP), Trieste 34151, Italy}
\email{ahassana@ictp.it}

\title[An \textsf{achemso} demo]
  {Investigating the Role of pH and Counterions in the Intrinsic Fluorescence of Solid-State \textup{\textsmaller{L}}-Lysine}

\begin{document}

\begin{abstract}
There is currently a growing interest in understanding the origins of intrinsic fluorescence as a way to design non-invasive probes for biophysical processes. In this regard, understanding how pH influences fluorescence in non-aromatic biomolecular assemblies is key to controlling their optical properties in realistic cellular conditions. Here, we combine experiments and theory to investigate the pH-dependent emission of solid-state \textup{\textsmaller{L}}-Lysine (Lys). Lys aggregates prepared at different pH values using HCl and H$_2$SO$_4$ exhibit protonation- and counterion-dependent morphology and fluorescence, as shown by microscopy and steady-state measurements. We find an enhancement in the fluorescence moving from acidic to basic conditions. To uncover the molecular origin of these trends, we performed non-adiabatic molecular dynamics simulations on three Lys crystal models representing distinct protonation states. Our simulations indicate that enhanced protonation under acidic conditions facilitates non-radiative decay via proton transfer, whereas basic conditions favor radiative decay. Our combined experimental-theoretical work highlights pH and counterion identity as key factors tuning fluorescence in Lys assemblies, offering insights for designing pH responsive optical materials based on non-aromatic amino acids.
\end{abstract}

\begin{tocentry}

\includegraphics[width=\linewidth]{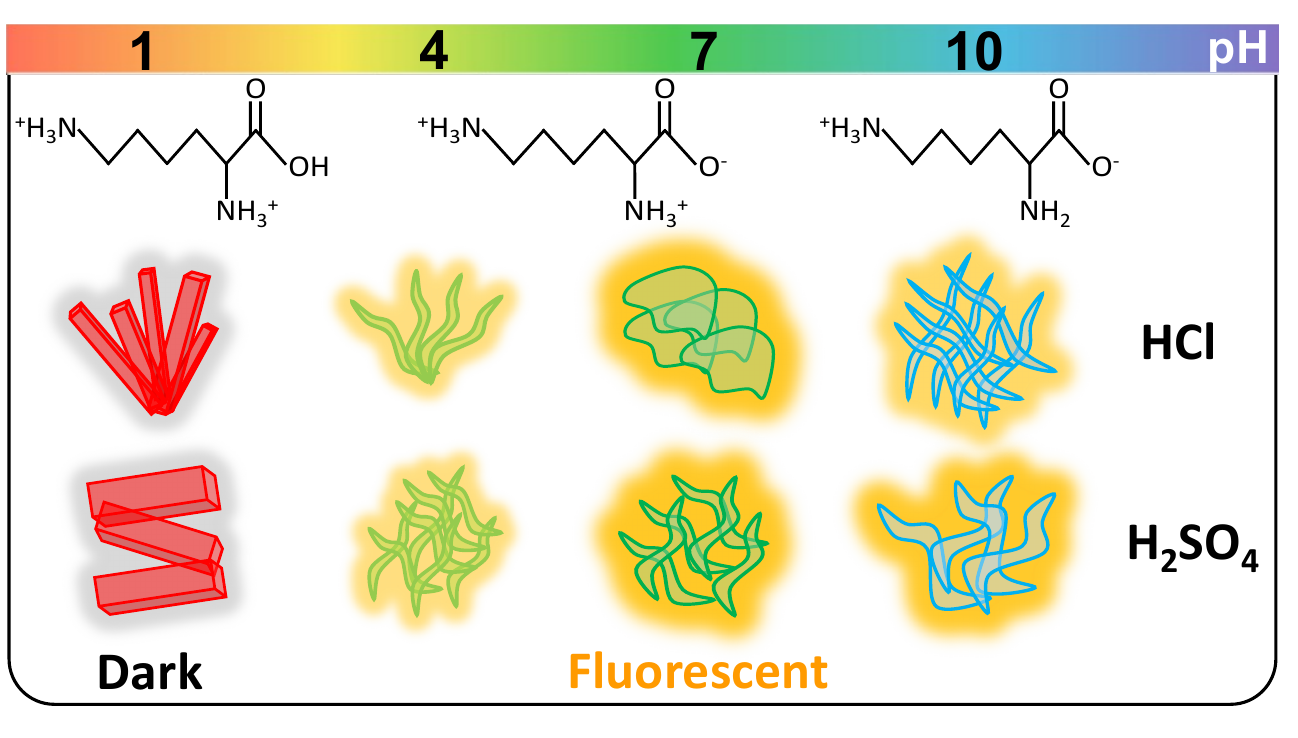}

\end{tocentry}

\section{Introduction}
Fluorescence in biological materials is typically attributed to aromatic amino acids due to their conjugated $\pi$-systems and favorable photophysical properties\cite{Lakowicz_2006}. However, recent studies have revealed that non-aromatic amino acids and peptides can also exhibit intrinsic photoluminescence, particularly under aggregation in solution or in the solid state.\cite{Pinotsi2016,Grisanti2020,arnon2021off,new_tang2021nonconventional,miron2023,Banerjee_2025,zhang2025_gluser} A key mechanism proposed to underlie this phenomenon is aggregation-induced emission (AIE),\cite{Luo_2001} wherein non-emissive species in dilute solution become strongly emissive upon aggregation. While AIE is well established in synthetic organic luminogens,\cite{He_2018,Zhao_2020,yu_2020} its manifestation in non-aromatic biomolecular assemblies, such as peptide and amino acid aggregates, has opened new directions for exploring photophysics in biologically relevant systems.\cite{delmercato_2007,new_chen2018prevalent,Lapshina_2019,Zhang_2019}

Hydrogen bonding plays a central role in mediating luminescence in these non-aromatic systems, influencing both the extent of electron delocalization and the vibrational modes that promote non-radiative decay from the excited state (ES)\cite{SHUKLA_2004,Pinotsi2016,Stephens2021,miron2023}. Environmental factors such as pH, solvent polarity, and ionic strength can profoundly affect these interactions, often altering morphology and photophysical properties\cite{Fung_2006,Clementi_2019}. However, how these factors reshape the ES potential energy landscape remains unclear\cite{Wang_2025}. While peptide self-assembly under neutral pH has been extensively studied,\cite{Wang_2018,Sarkhel_2020,McCahill_2024} less attention has been given to extreme pH conditions, which can drastically alter conformation, charge distribution, and intermolecular interactions\cite{Li_2022,Wang_2025}. In this regard, pH is known to play an important role in tuning fluorescence in aromatic or more generally conjugated molecular systems\cite{white1959effect,STEINER1963341,berezin2009ph,Han_2010}. Thus, the potential of designing novel probes based on intrinsic fluorescence as indicators of pH in optical sensing and imaging, could open up new applications in biological and material sciences\cite{Han_2010,Steinegger_2020}. 

\begin{figure}[h!]
    \centering
    \includegraphics[width=\linewidth]{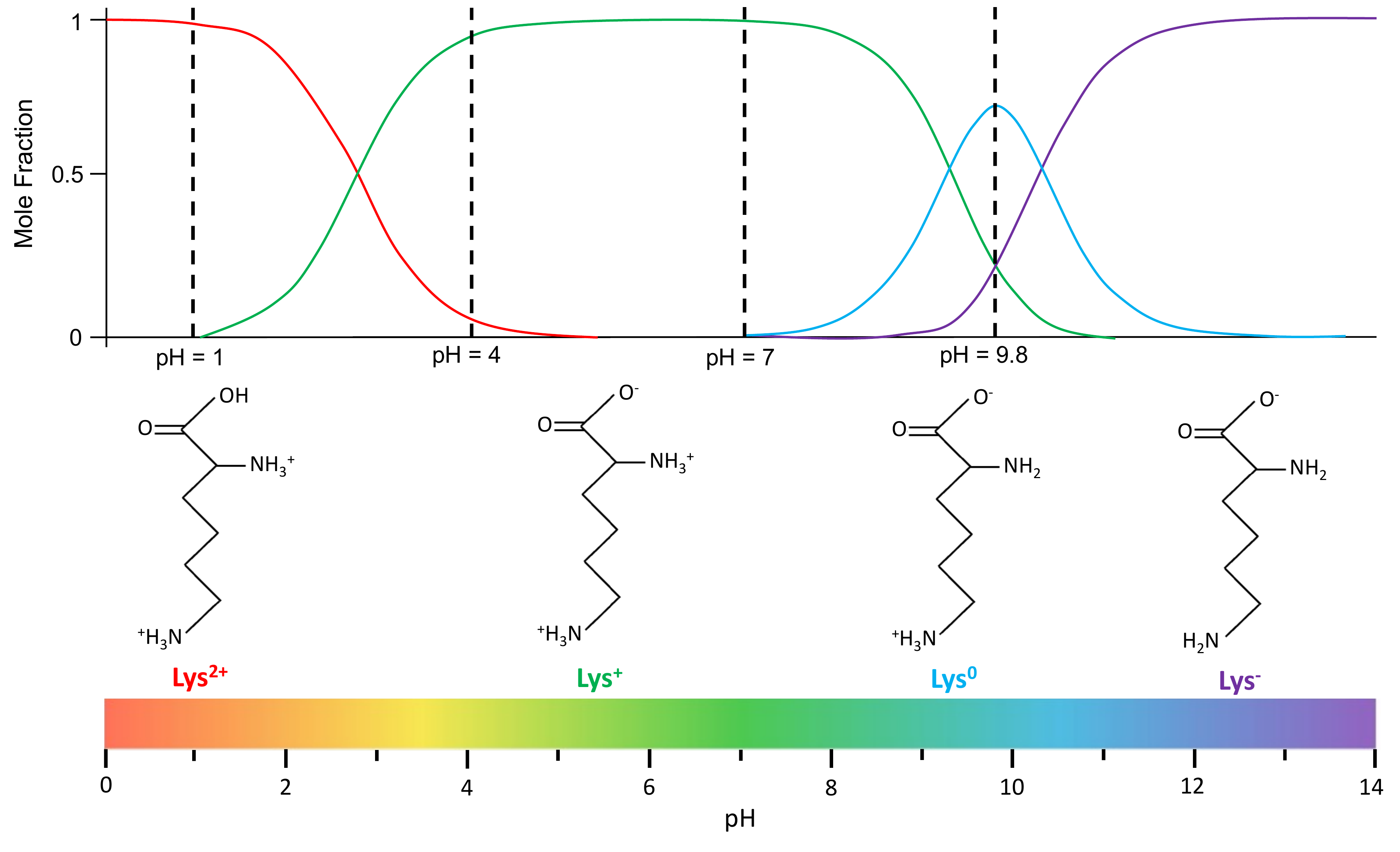}
    \caption{Mole fractions of the four protonation states of \textup{\textsmaller{L}}-lysine as a function of pH. The fully protonated cationic species (Lys$^{2+}$) and fully deprotonated anionic species (Lys$^{-}$) each reach near-complete population under highly acidic or basic conditions, respectively. The zwitterionic form (Lys$^0$) peaks at approximately 75$\%$ near the isoelectric point (pH 9.74), reflecting the similar pK$_\text{a}$ values of the $\alpha$- and $\varepsilon$-amino groups.}
    \label{fig:Fig1}
\end{figure}

In this context, \textup{\textsmaller{L}}-Lysine (Lys), a non-aromatic amino acid with a protonatable side chain, provides a compelling model to probe pH-dependent luminescence. Depending on pH, Lys adopts distinct protonation states (see Fig. \ref{fig:Fig1}),\cite{Nolting_2007} which in turn strongly influence its self-assembly and crystal packing, thereby modulating its photophysical properties.\cite{stagi2022,arnon2021off}. Similar effects have been observed in poly-lysine systems, where pH influences aggregate morphology and optical response.\cite{Stagi_2021,stagi_2022_polyLys,Kastinen_2023_polyLys}. Recent work by Stagi \textit{et al.}\cite{stagi2022} combined experiments and theory to investigate the photophysical properties of self-assembled Lys in aqueous solution. Absorption spectra showed a distinct band at 278 nm, attributed to electronic transitions in hydrogen bonded Lys clusters. Classical molecular dynamics (MD) simulations further confirmed that Lys forms clusters of varied size through hydrogen bonds (HBs) and hydrophobic interactions, with the former playing a dominant role. Fluorescence measurements in solution exhibit strong pH dependence: emission is enhanced under neutral and basic conditions but quenched at low pH, where high protonation is proposed to inhibit functional group interactions, reducing emission efficiency.\cite{stagi2022} However, the precise ES relaxation pathways that affect fluorescence emission in titratable amino acids, such as lysine, remains poorly understood.

In this work, by combining both experiments and theory, we address a gap in our understanding of how pH affects non-aromatic fluorescence using solid-state lysine as a model system. Solid-state Lys aggregates were prepared at four pH values (1, 4, 7, and 10) to probe how the protonation influences both morphological and photoluminescence properties. Since pH was adjusted with either a monoprotic (HCl) or a diprotic (H$_2$SO$_4$) acid, we also assessed the impact of the counterion on these properties. Morphology was characterized by optical and scanning electron microscopy (SEM), while photoluminescence was studied via confocal microscopy, steady-state fluorescence, and external quantum efficiency (EQE) measurements. Our results show that both protonation state, the choice of the acid and subsequently the counterion in the corresponding crystal, significantly influence aggregate morphology and fluorescence, establishing these as key parameters for tuning optical properties. To uncover the molecular origins behind these trends, we performed non-adiabatic molecular dynamics (NAMD) simulations to follow the ES dynamics of Lys. While direct simulation of the experimental aggregates is precluded by the lack of precise structural data, we can instead model crystal structures of Lys representing distinct protonation states (Lys$^{2+}$, Lys$^{+}$, and Lys$^0$ in Fig. \ref{fig:Fig1}).\cite{Capasso_1983, wright_1962, Williams_2015} These models offer well-defined frameworks to interpret fluorescence behavior and examine ES processes. Our simulations indicate that fluorescence decreases from low to high protonation states ($\text{Lys}^0 \gtrsim \text{Lys}^{+} > \text{Lys}^{2+}$), consistent with our experimental data and prior reports\cite{stagi2022,arnon2021off}. In short, increased protonation promotes proton transfer (PT) processes that enable non-radiative decay, causing fluorescence quenching under acidic conditions. 

\section{Methods}
\subsection{Experimental Details}

\subsubsection{Chemicals and Samples Preparation}

\textup{\textsmaller{L}}-Lysine (Lys) monohydrate, HCl ($37\%$wt) and H$_2$SO$_4$ ($96\%$wt) were purchased by Thermo Fisher Scientific (Milan, Italy). All solutions were prepared by weight dissolving Lys in doubly distilled water; their pH was adjusted using HCl ($37\%$wt) and H$_2$SO$_4$ ($96\%$wt). pH was measured by pH-meter (Mettler Toledo FiveEasy Plus pH meter FP20, equipped with a micro electrode). Experimental conditions were verified to be above the critical aggregation concentration (CAC) of Lys, determined through fluorescence probe assays, as reported in details in the Supporting Information (SI).

\subsubsection{Scanning Electron Microscopy}

Morphological characterization of the samples was performed using field-emission scanning electron microscopy (ThermoFisher Scientific Phenom Pro G6). Aliquots of 10 $\mu$L Lys solutions were drop-cast onto aluminum stubs and air-dried at room temperature. Samples were sputter-coated with a 7 nm gold layer (25 mA, 75 s) prior to imaging. Images were acquired in high vacuum mode at an accelerating voltage of 15 kV using the Secondary Electron Detector.

\subsubsection{Optical Microscopy}

Samples were prepared by drop-casting 20 $\mu$L of each Lys solution onto glass microscope slides, which were then air-dried overnight at room temperature. The dried samples were examined using an Optech BM80Pol microscope (Milan, Italy) under both cross-polarized and bright-field light.

\subsubsection{Confocal Microscopy}

Samples of Lys solutions at various protonation states were drop-cast onto microscope slides and allowed to air-dry overnight. The dried samples were analyzed on the MICA microhub (Leica, Wetzlar, Germany) platform using $\times1.6$ and $\times10$ magnifications. Fluorescence images were acquired using filters in the spectral regions of blue ($\lambda_{exc} = 402~\text{nm}$, $\lambda_{em} = 421 ~\text{nm}$), green ($\lambda_{exc} = 495~\text{nm}$, $\lambda_{em} = 519~\text{nm}$), and red ($\lambda_{exc} = 590~\text{nm}$, $\lambda_{em} = 617~\text{nm}$).

\subsubsection{Fluorescence Spectroscopy}

Fluorescence spectra were acquired at room temperature on a Jasco spectrofluorophotometer (Model FP-8550), equipped with a continuous output Xenon arc lamp. Measurements in solution were conducted in a quartz cell with a 1.0 cm path length. Solid state fluorescence and photoluminescence quantum yields (PLQYs) were evaluated using an ILF-835 integrating sphere. The absolute method was employed for PLQY determination thereby overcoming the issue of unavoidable differences in the relative emission intensities arising from poorly reproducible drop-casted amino acid clusters. Fluorescence spectra of solid-state Lys were collected up to 600 nm by scanning the 330–410 nm excitation range in 10 nm increments. All acquisitions were averaged over five measurements after spectral correction. 

\subsection{Computational Details}

\subsubsection{Classical Molecular Dynamics}

Lys systems in different protonation states were modeled using crystallographic structures of \textup{\textsmaller{L}}-Lysine sulfate (LLS), \textup{\textsmaller{L}}-Lysine monohydrochloride (LLMHCl), and \textup{\textsmaller{L}}-Lysine hemihydrate (LLH). Supercells of appropriate size were built for each system: 15$\times$8$\times$8 for LLS, 11$\times$5$\times$11 for LLMHCl, and 7$\times$11$\times$6 for LLH. Classical molecular dynamics (MD) simulations were performed using the AMBER MD package version 2023\cite{amber2005,ambertools2023} with the Generalized Amber Force Field (GAFF) \cite{gaff_wang2004} applied to all molecular mechanics (MM) atoms. 
For each supercell, we first performed energy minimization to remove any unfavorable contacts. Minimized systems were then equilibrated for 10 ns in the NVT ensemble at 300 K (2 fs time step, Langevin thermostat\cite{Izaguirre_2001} with $\gamma=1.0$ ps$^{-1}$). Root mean square deviation (RMSD) relative to the first frame and potential energy were monitored along the simulations, as shown in Fig. \ref{fig:SI_1} of the SI, confirming that the systems equilibrate within 10 ns. For each system, 50 independent frames were extracted from the corresponding MD production and used as starting geometries for the subsequent quantum mechanics/molecular mechanics (QM/MM) simulations\cite{qmmm_warshel_levitt_1976,qmmm_brunk2015mixed}.

\subsubsection{Ground State QM/MM}

For each crystal, we selected a suitable QM region (see Fig. \ref{fig:Fig5}a–c) which was then equilibrated at 300 K using ORCA 5.0.4\cite{orca_neese2012,orca_neese2018,orca_neese2020,orca_neese2022,orca_neese2023shark} (QM) interfaced with AMBER 2023\cite{amber2005,ambertools2023} (MM). The QM region was treated using the CAM-B3LYP exchange-correlation functional\cite{camb3lyp_yanai2004} and the def2-SVP basis set\cite{def2_weigend2005,def2_hellweg2015} with the RIJ-COSX approximation\cite{rijcosx_neese2009} to accelerate Coulomb and exact exchange integral evaluations. A first equilibration of 1 ps was performed in the NVT ensemble with tight thermostat coupling ($\gamma=10.0$ ps$^{-1}$, 0.5 fs step, no bond constraints), followed by 1 ps with a standard thermostat coupling ($\gamma=1.0$ ps$^{-1}$). Final nuclear coordinates and velocities were used as initial conditions for the non-adiabatic molecular dynamics (NAMD) simulations described below. 

The thermally equilibrated structures were further used to compute vertical excitation energies at the time-dependent density functional theory (TDDFT) level. These calculations were performed using the same ORCA/AMBER QM/MM framework described above, but with no MD propagation (0 MD steps). For each Lys crystal system, the ten lowest singlet excited states were evaluated to extract the absorption spectra of the different crystal structures. Molecular orbitals were computed using a post-processing tool available in ORCA.

\subsubsection{Non-adiabatic Molecular Dynamics}

NAMD simulations were performed using the Trajectory Surface Hopping (TSH) method within the SHARC formalism\cite{sharc_richter2011,sharc_mai2018}, which implements Tully's Fewest Switches Surface Hopping (FSSH)\cite{tsh_tully1990molecular} and allows the inclusion of time-dependent non-adiabatic couplings (TD-NACs). In this approach, nuclei evolve classically on electronic potential energy surfaces computed on-the-fly. For a comprehensive overview of TSH methodology, the reader is referred to Refs. \cite{tsh_barbatti2011nonadiabatic,malhado2014non,persico2014overview,tsh_crespo2018recent}. 

For each system, 50 trajectories were propagated including the first five singlet states ($\text{S}_0-\text{S}_4$) to adequately describe the lowest excited states relevant for fluorescence studies. All trajectories were initialized on the first excited state ($\text{S}_1$) and propagated in the NVE ensemble for 500 fs with 0.5 fs time step using the velocity Verlet integration algorithm\cite{verlet1967computer,swope1982computer}. The ground-state QM/MM functional and basis set were retained. Excitation energies and oscillator strengths were calculated at the TDDFT\cite{tddft_casida1995} level within the Tamm-Dancoff approximation (TDA)\cite{tda_hirata1999}, which improves TDDFT behavior near conical intersections\cite{Cordova_2007,Tapavicza_2008,Peach_2012}. NACs were computed via the Hammes-Schiffer–Tully scheme\cite{tsh_hammes1994proton} based on wavefunction overlap integrals between successive time steps. Electronic dynamics were integrated using 25 sub-steps per nuclear time step, and decoherence effects were included using the energy-based correction of Granucci \textit{et al.}\cite{tsh_granucci2007critical} with a 0.1 Hartree parameter. Surface hops to S$_0$ were allowed when the energy gap between the current state and S$_0$ was below 0.2 eV, consistent with prior TDDFT-based NAMD studies.\cite{plasser2014surface,wen2023excited,ibele2020excimer} Kinetic energy was rescaled along the direction of the NAC vector to conserve energy during allowed hops; frustrated hops were discarded. Simulations were performed using SHARC 3.0.1\cite{SHARC3.0} for surface hopping, ORCA 5.0.4\cite{orca_neese2012,orca_neese2018,orca_neese2020,orca_neese2022,orca_neese2023shark} for quantum chemical calculations, and TINKER 6.3.3\cite{tinker_ponder2004} for MM dynamics. 

\section{Results and Discussion}

\subsection{Experimental Results}

\subsubsection{Structural Characterization of Lysine Aggregates}

We begin by examining the morphology of solid-state Lys aggregates prepared at pH 1, 4, 7, and 10, with pH adjusted using either HCl or H$_2$SO$_4$, by optical and scanning electron microscopy. The characterization was performed on Lys samples prepared in water at the concentration of $1$ g/mL ($6.8$ mol/L) corresponding to the maximum of its solubility. At this concentration, the experimentally measured pH is 11.8, approximately two units above the isoelectric point (9.74.) Under these conditions, the concentration is approximately tenfold higher than the CAC, ensuring the formation of aggregated clusters (see Fig. \ref{fig:SI_2} and discussion in the SI).

\begin{figure}[h!]
    \centering
    \includegraphics[width=\linewidth]{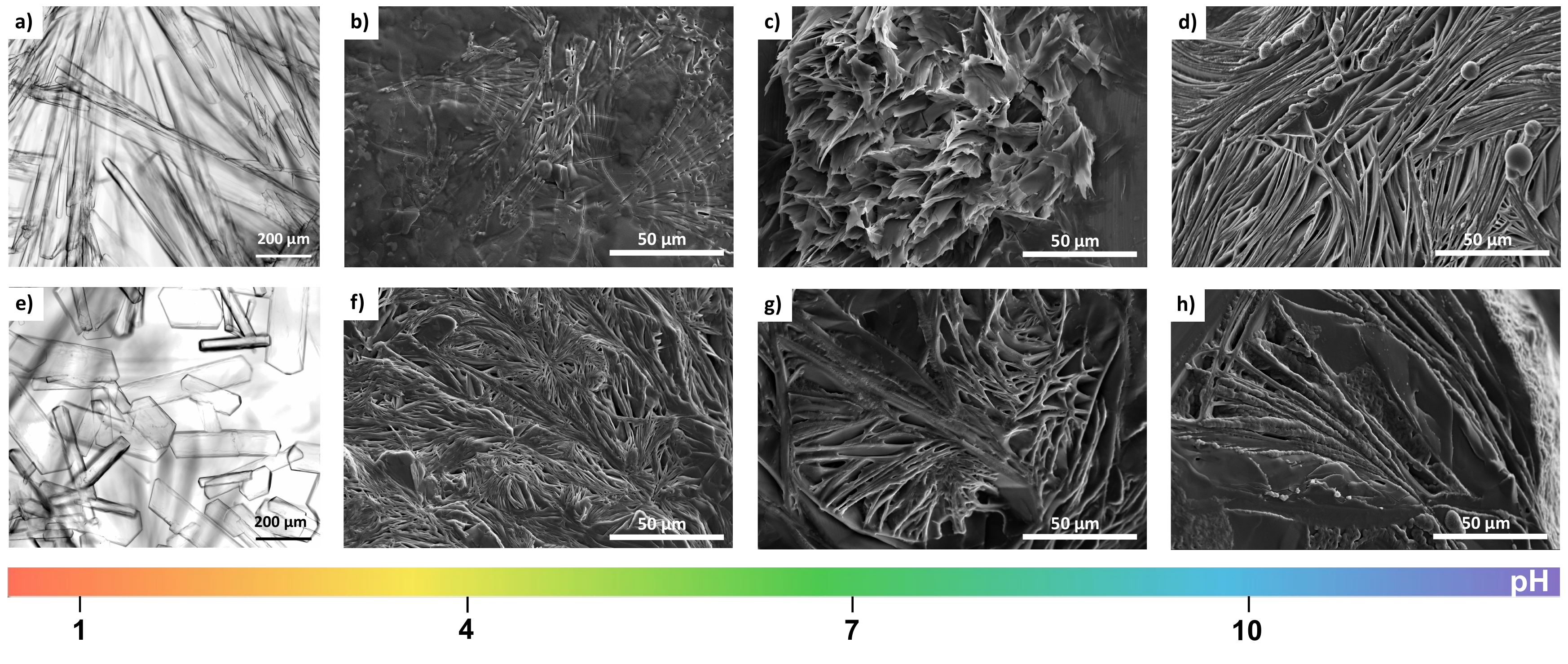}
    \caption{Experimental morphological characterization of solid-state \textup{\textsmaller{L}}-Lysine (Lys) aggregates prepared at pH 1, 4, 7, and 10 using HCl (panels a–d) and H$_2$SO$_4$ (panels e–h) as acidifying agents. (a, e) Optical microscopy images of crystalline Lys aggregates formed by drop-casting 6.8 mol/L solutions adjusted to pH 1. (b–d, f–h) Scanning electron microscopy (SEM) images of Lys aggregates at pH 4, 7, and 10 for HCl (b–d) and H$_2$SO$_4$ (f–h) conditions. All SEM images were acquired at 3000x magnification with a scale bar of 50 $\mu$m.}
    \label{fig:Fig2}
\end{figure}

Fig. \ref{fig:Fig2} shows clear morphological differences depending on both acid type and pH. Under strongly acidic conditions (pH = 1), both HCl- and H$_2$SO$_4$-treated samples initially formed oily films at room temperature, with no macroscopically ordered structures observed over time. This behavior can be rationalized by considering that, at this pH, Lys is in its double protonated state, which introduces strong electrostatic repulsion between molecules. The presence of counterions (Cl$^-$ or SO$_4^{2-}$) can screen these interactions, enabling molecular association. However, at room temperature, the combination of high solubility,\cite{Amend_1997} thermal motion, and residual repulsion may hinder the establishment of long-range order. Upon overnight cooling to $4^\circ\text{C}$, reduced molecular motion enhances the stability of weak directional interactions, leading to the formation of well-ordered crystals (Fig. 2a, Lys + HCl; Fig. 2e, Lys + H$_2$SO$_4$), in line with previous studies on cooling-induced crystallization\cite{Christopher_1998}. Optical microscopy revealed distinct morphologies: HCl-treated samples formed thinner, densely packed crystals, whereas H$_2$SO$_4$ promoted larger, plate-like ones, suggesting that the two counterions influence nucleation behavior differently. Reheating caused crystals to revert to oily films, precluding further morphological study, highlighting the role of temperature in tuning the equilibrium between disordered and crystalline phases.\cite{Chen_2022,Monti_2023}.

At pH 4, HCl-treated Lys aggregates (Fig. \ref{fig:Fig2}b) formed thin and disperse small fibers, whereas H$_2$SO$_4$-treated aggregates (Fig. \ref{fig:Fig2}f) resulted in denser fibrous networks with a more compact appearance. This contrast suggests that sulfate ions can promote stronger cross-linking, leading to a more interconnected aggregation. At pH 7, HCl-derived Lys aggregates (Fig. \ref{fig:Fig2}c) formed a layered, sheet-like morphology, while those acidified with H$_2$SO$_4$ (Fig. \ref{fig:Fig2}g) a more porous, interconnected network, suggesting that sulfate anions promote a web-like assembly rather than compact layers. At pH 10, both acids yielded fibrous architectures (Fig. \ref{fig:Fig2}d, and Fig. \ref{fig:Fig2}h), with thinner and more interconnected fibers in HCl-treated samples. Cross-polarized light microscopy (Fig. \ref{fig:SI_3}) revealed birefringence, indicating isotropic organization of the fibrillar network at this pH. 

\subsubsection{pH-Dependent Fluorescence in Lysine Aggregates}

Since both pH and acid type influence Lys aggregate morphology, and aggregate structure is known to affect the emergence and intensity of intrinsic fluorescence in non-aromatic systems,\cite{berger_2015,Pinotsi2016,stagi2022} we next investigated how these morphological variations affect the optical response. To establish this, we turned to fluorescence-based techniques, starting with confocal and fluorescence microscopy, followed by steady-state spectroscopy and quantum efficiency measurements to quantify the pH-dependent emission.

\begin{figure}[h!]
    \centering
    \includegraphics[width=\linewidth]{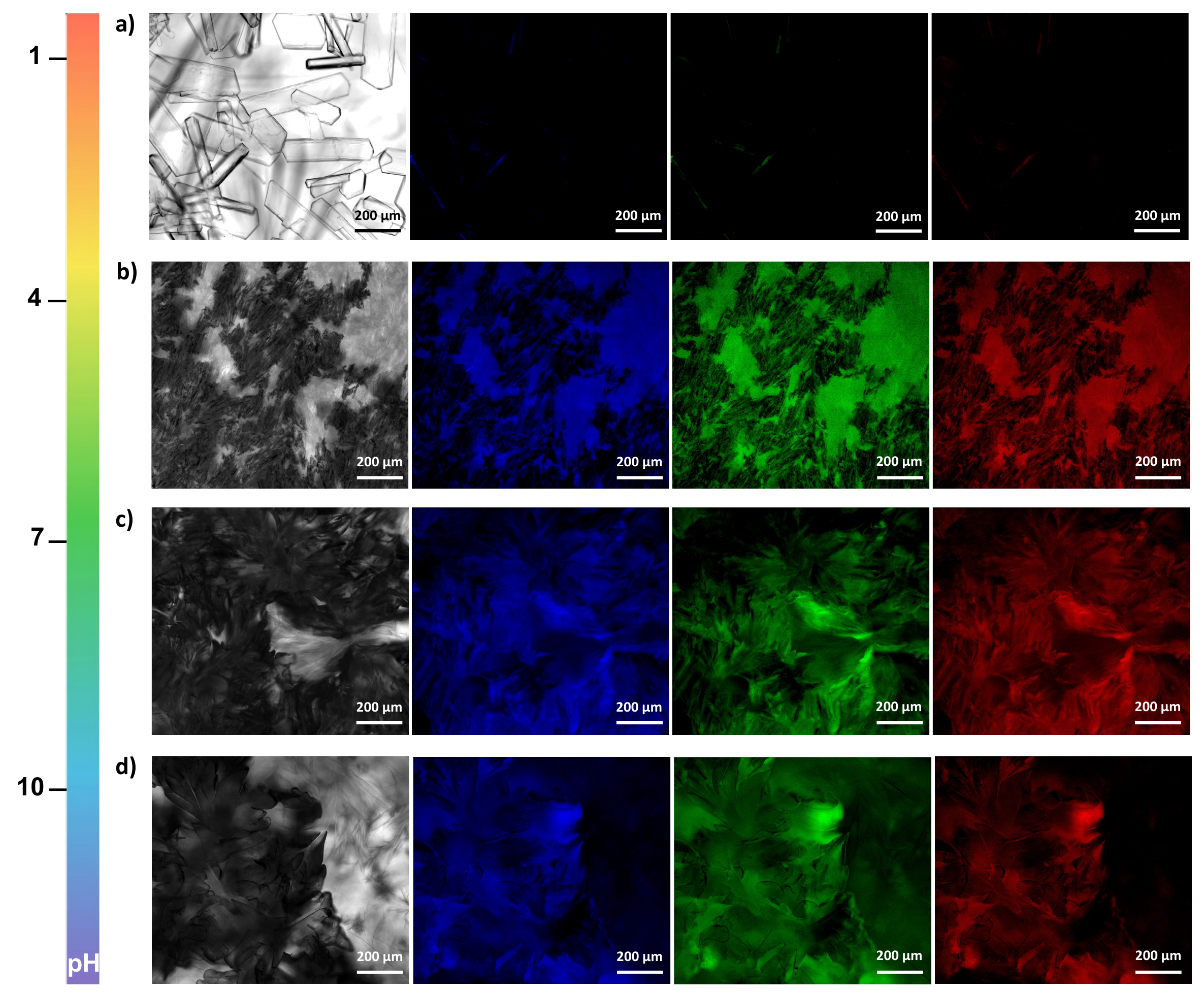}
    \caption{Confocal and fluorescence microscopy of solid-state \textup{\textsmaller{L}}-Lysine (Lys) aggregates prepared at various pH values. Aggregates were deposited on clean coverslip glass and air-dried at room temperature. Confocal microscopy images of Lys aggregates formed at pH 1, 4, 7, and 10 using H$_2$SO$_4$ as the acidifying agent are shown in panel a, b, c, and d, respectively. For each pH, images are arranged left to right as follows: panel a begins with an optical microscopy image (dark-field data unavailable), followed by fluorescence images in the blue ($\lambda{\text{exc}}$ = 402 nm, $\lambda{\text{emi}}$ = 421 nm), green ($\lambda_{\text{exc}}$ = 495 nm, $\lambda_{\text{emi}}$ = 519 nm), and red ($\lambda_{\text{exc}}$ = 590 nm, $\lambda_{\text{emi}}$ = 617 nm) channels; panels (b–d) begin with dark-field images followed by the same fluorescence channels.}
    \label{fig:Fig3}
\end{figure}

Confocal microscopy of Lys aggregates prepared with HCl (Fig. \ref{fig:SI_4}) and H$_2$SO$_4$ (Fig. \ref{fig:Fig3}) confirmed the evolution from crystalline‑like structures at pH 1 to diffuse fibrillar networks at pH 4, 7, and 10. This trend is evident in the far-left subpanels of Fig. \ref{fig:Fig3}b–d (dark-field images). For pH 1 (Fig \ref{fig:Fig3}a), optical microscopy is shown due to the absence of dark-field data. Each panel also includes fluorescence microscopy images recorded in the blue (second column), green (third column), and red (rightmost column) channels. A striking enhancement in fluorescence intensity is observed with increasing pH for both acids: aggregates at pH 1 exhibit weak or negligible emission across all channels, while those at neutral and basic conditions (Fig. \ref{fig:Fig3}b–d) show stronger fluorescence. 

To quantify this, we collected steady-state fluorescence spectra across excitation wavelengths from 330-410 nm (10 nm step). The resulting spectra for Lys acidified with HCl and H$_2$SO$_4$ are shown in panels a-d and panels e-h of Fig. \ref{fig:SI_5}, respectively. A linear correlation between excitation and emission wavelengths was observed (Fig. \ref{fig:SI_6}), consistent with previous reports on Lys aggregates in solution\cite{stagi2022}, and poly-lysine polymers\cite{Stagi_2021}, where this behavior was attributed to the ability of Lys to form aggregates of different size. Such excitation-dependent fluorescence, or red-edge excitation shift, has been observed in a variety of systems, including carbon nanodots\cite{Sharma_2016}, sodium-uracil thin films\cite{pale_2017}, self-assembled peptide nucleic acids\cite{berger_2015}, and organic fluorescent nanoparticles\cite{lee2013}. Proposed origins include heterogeneous aggregate populations\cite{Sharma_2016}, slow relaxation of the local environment around the excited fluorophore\cite{berger_2015,pale_2017}, and collective proton motions within HB networks that modulate the ES manifold\cite{lee2013}. We propose that, in our Lys samples, the excitation-dependent emission likely arises from similarly heterogeneous supramolecular environments shaped by protonation state and ionic interactions.

\begin{figure}
    \centering
    \includegraphics[width=0.55\textwidth]{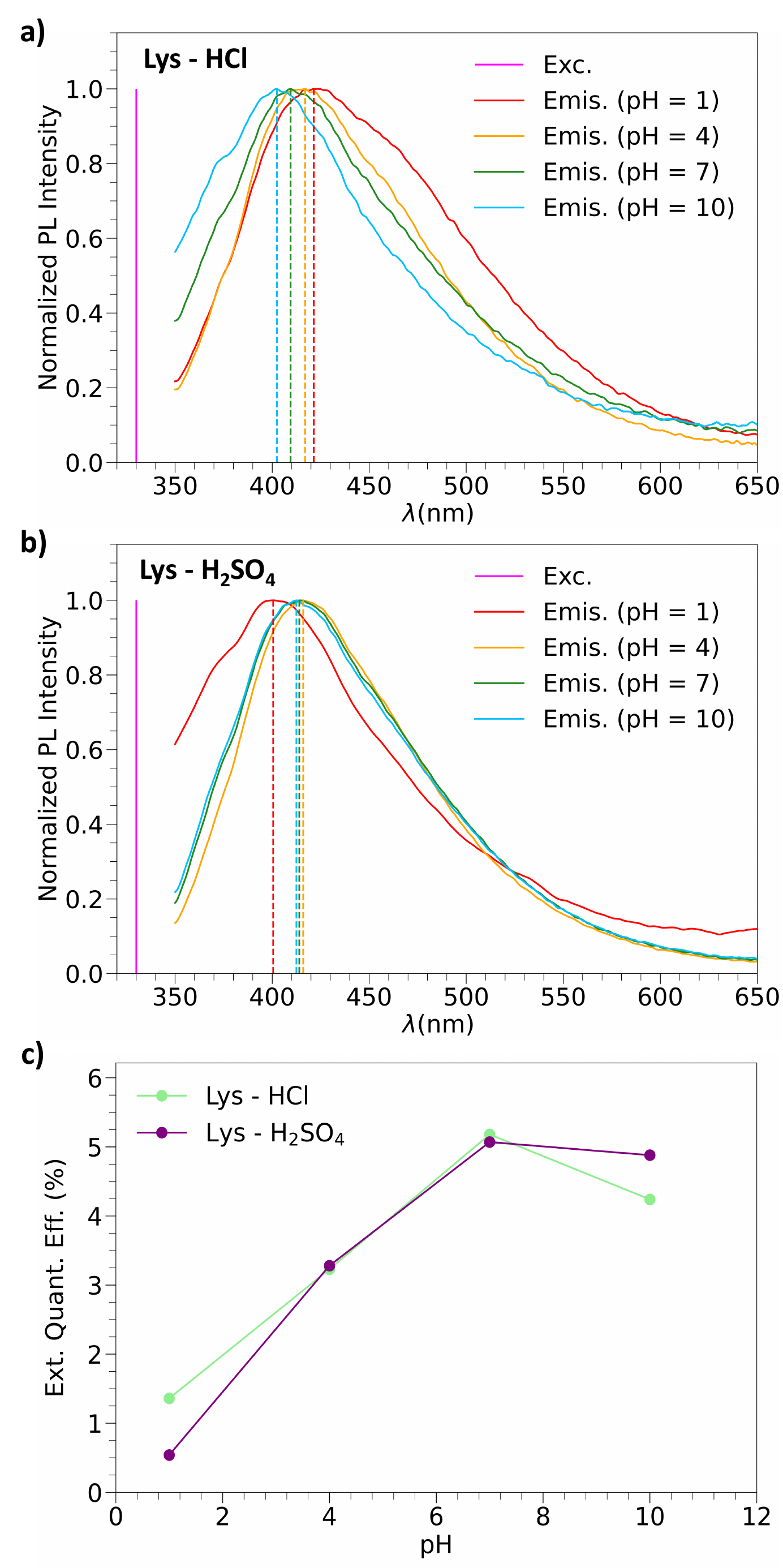}
    \caption{Quantitative optical characterization of solid-state \textup{\textsmaller{L}}-Lysine (Lys) aggregates prepared at different pH values. Panels a and b show normalized steady-state fluorescence spectra ($\lambda_\text{exc}$ = 330 nm, magenta vertical line) of solid Lys samples prepared with HCl and H$_2$SO$_4$, respectively. Emission spectra are displayed in red, orange, green, and cyan for pH 1, 4, 7, and 10. Corresponding vertical dashed lines in the same colors indicate the emission maxima. Panel c shows the corresponding external quantum efficiency (Ext. Quant. Eff.) measurements: green for HCl and purple for H$_2$SO$_4$.}
    \label{fig:Fig4}
\end{figure}

To isolate the role of protonation and acid identity, we compared normalized fluorescence spectra at $\lambda_\text{exc} = 330$ nm for all samples, as shown in Fig. \ref{fig:Fig4}a-b. Both acid treatments yielded broad emissions (350–550 nm), with emission maxima ($\lambda_{max}$) centered between 400 nm and 420 nm, depending on pH and acid. For HCl (Fig. \ref{fig:Fig4}a), $\lambda_{\text{max}}$ blue-shifts from 422 nm (pH 1, red line) to 403 nm (pH 10, cyan line), while for H$_2$SO$_4$ (Fig. \ref{fig:Fig4}b), the pH 1 sample shows the most blue-shifted emission ($\lambda_{max} = 401$ nm), and all other pH values exhibit nearly identical spectra ($\lambda_{max}\sim$ 414 nm). These subtle spectral differences qualitatively correlate with the SEM observations. H$_2$SO$_4$-treated samples exhibit more homogeneous fibrous morphologies across pH, consistent with a uniform molecular organization and emission response. In contrast, the greater morphological variability observed for HCl-treated samples suggests a broader range of local packing arrangements, which may contribute to the observed spectral differences. Although this remains an empirical correlation, the SEM data provide qualitative insight into the molecular arrangement influencing fluorescence behavior, in line with previous reports correlating aggregate morphology and emission properties\cite{Hong_2011}.

While these spectra offer an initial view of how pH influences fluorescence, their normalization masks variations in absolute emission intensity. To address this, we measured external quantum efficiency (Fig. \ref{fig:Fig4}c), which accounts for light extraction efficiency\cite{Park_2024} and is therefore particularly relevant for assessing emission in solid-state devices\cite{liu_2017,song_2018, Jang_2020}. A clear pH dependence emerges: EQE increases from $\sim0.5-1\%$ at pH 1 for both acid treatments to nearly 5$\%$ at pH 7 for Lys with HCl (green plot) and at pH 7-10 for Lys with H$_2$SO$_4$ (purple plot), reflecting higher emission efficiency under low protonation. These trends are consistent with previous studies on solution-phase Lys aggregates, which linked protonation to fluorescence quenching\cite{stagi2022}. Notably, the two acids impart distinct efficiencies: with H$_2$SO$_4$, EQE increases from pH 1 to pH 7 and then plateaus at pH 10, whereas with HCl the efficiency peaks at pH 7 and declines at pH 10. This divergence underscores the additional role of the counterion, beyond protonation alone, in dictating aggregate structure and, consequently, solid‑state fluorescence efficiency. 

\subsection{Theoretical Results}

\subsubsection{Excited State Simulations of Lysine Crystals}

\begin{figure}[h!]
    \centering
    \includegraphics[width=\linewidth]{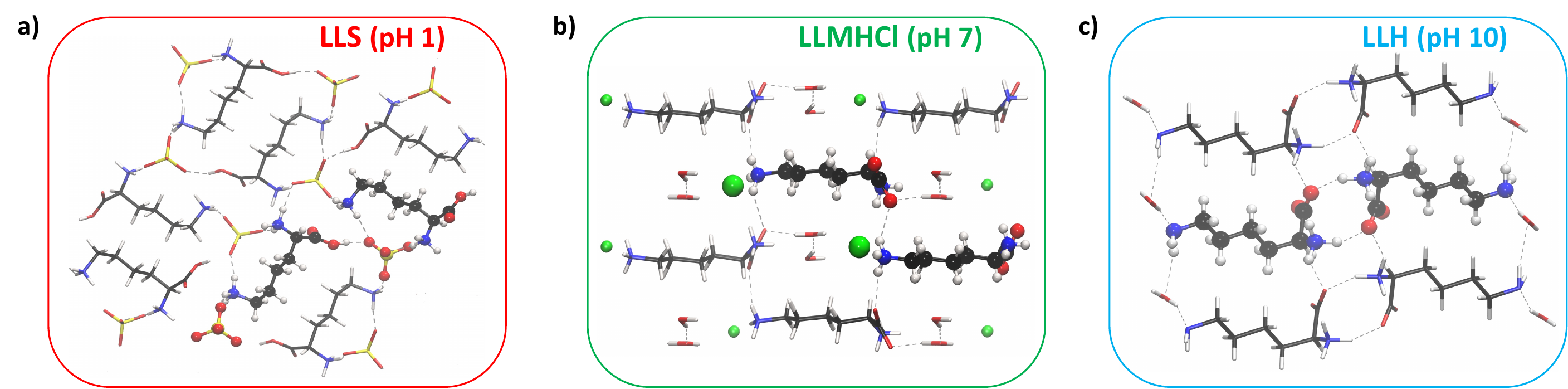}
    \caption{Crystal structures used to model the different protonation states of \textup{\textsmaller{L}}-lysine: (a) \textup{\textsmaller{L}}-lysine sulfate (LLS), (b) \textup{\textsmaller{L}}-lysine monohydrochloride dihydrate (LLMHCl), and (c) \textup{\textsmaller{L}}-lysine hemihydrate (LLH).}
    \label{fig:Fig5}
\end{figure}

Our experimental characterization shows that both pH and counterion modulate the emission of solid‑state Lys aggregates, implicating changes in protonation and aggregation as key factors. However, experimental techniques alone cannot directly resolve the molecular origins of this effect, specifically, how protonation state shapes the electronic structure and ES dynamics. To address this, we performed NAMD simulations on three Lys crystal models representing distinct protonation states, as shown in Fig. \ref{fig:Fig5}. 

The structural differences among the lysine crystals reflect protonation-dependent variations that mirror the pH effects observed in aggregate morphologies, supporting their use as proxies to study solid-state Lys photophysics. As we will see shortly, these models also correctly reproduce the trends in the emission observed as a function of pH. \textup{\textsmaller{L}}-Lys sulfate (LLS, Fig. \ref{fig:Fig5}a) adopts a folded conformation stabilized by a dense HB network between SO$_4^{2-}$ and the -NH$_3^+$ and CO$_2$H groups of lysine\cite{Capasso_1983,KRISHNAKUMAR_2010}, with HBs exclusively between Lys and sulfate and no direct Lys-Lys interactions due to electrostatic repulsion. In \textup{\textsmaller{L}}-Lys monohydrochloride dihydrate (LLMHCl, Fig. \ref{fig:Fig5}b), all amines are protonated, the C-terminus is deprotonated, and both Lys–Lys and Lys–water HBs are observed. Cl$^-$ counterions further stabilize the structure via electrostatic interactions with both the $\alpha$- and $\varepsilon$-amino groups\cite{wright_1962,ramesh2006growth}. Finally, when \textup{\textsmaller{L}}-Lys crystallizes in its zwitterionic state, it forms a hemihydrate crystal structure (LLH, Fig. 5c). The Lys molecules orient such that the head groups--namely, the amino (-NH$_3^{+}$) and carboxyl (-CO$_2^-$) groups--of two monomers interact, while the side chain amino groups form HBs with water molecules, resulting in an extended HB network that propagates across the crystal layers\cite{Williams_2015,Williams_2016}. Interestingly, Arnon \textit{et al.} \cite{arnon2021off} showed that LLH crystals exhibit fluorescence, with its hydrogen bonding network likely playing a key role in the observed optical properties. However, their work did not include atomistic simulations to probe the mechanistic origin of the emission, which we address in the present study.

It is worth noting that, while multiple protonation states of lysine coexist at a given pH (see Fig. 1), each solid-state polymorph corresponds to a specific dominant protonation state as determined crystallographically\cite{wright_1962,Capasso_1983,Williams_2015,Williams_2016}. The protonation states present in the selected crystals (LLS, LLMHCl, and LLH) coincide with the maxima of the mole fraction distribution at pH 1, 4-7, and 10, respectively, making them natural proxies for our study. We acknowledge that this modeling strategy simplifies the complexity of real samples, where mixed protonation states and structural disorder may occur, but it allows us to isolate the effect of each dominant protonation state on fluorescence and to establish clear mechanistic connections between protonation, hydrogen-bonding networks, and excited state dynamics.

For each crystal, we performed 500 fs NAMD simulations, within a QM/MM framework, on 50 statistically independent initial conditions (see \textit{Computational Details}). In Fig. \ref{fig:Fig5}a–c, the lysine dimers rendered with glossy CPK spheres represent the QM regions used in our simulations. After 500 fs, most trajectories remained in the first ES for LLH (pH = 10) and LLMHCl (pH = 7), namely 98$\%$ and 92$\%$, whereas only 28$\%$ remained in S$_1$ for LLS (pH = 1), as shown in Fig. 6a (orange plot). The same panel shows a qualitative comparison between these fractions and the EQE values at different pH, providing an initial validation of our computational results. The consistency between the EQE and N$_{\text{S}1}$/N$_{\text{tot}}$ trends across pH supports the accuracy of our ES dynamics. Moreover, the severe quenching observed both in the experimental and calculated results at low pH highlights the key role of protonation in modulating ES lifetimes.

\begin{figure}[h!]
    \centering
    \includegraphics[width=\linewidth]{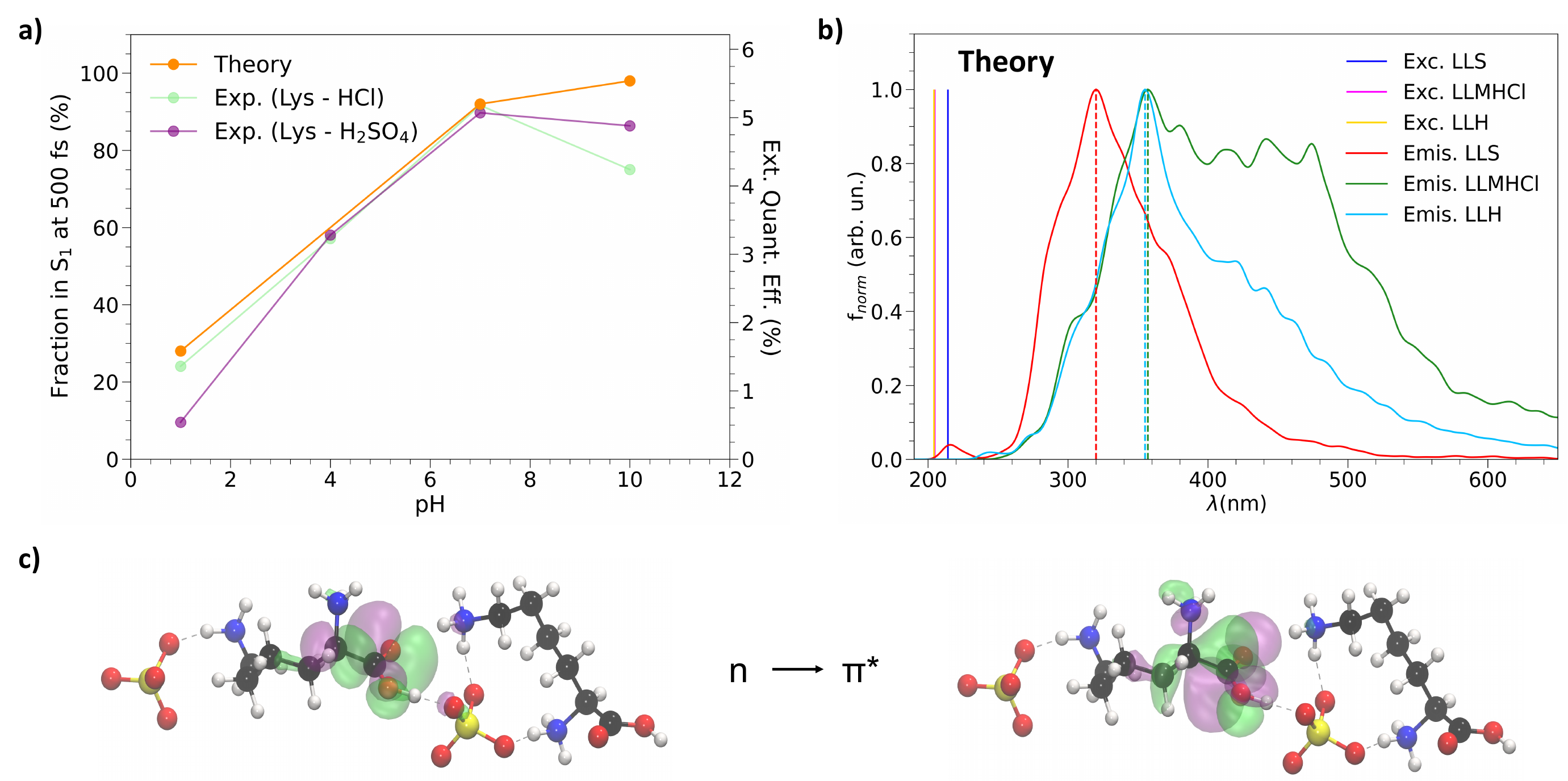}
    \caption{Panel a shows the fraction ($\%$) of surface-hopping trajectories remaining in the first excited state S$1$ (N$_{\text{S}1}$/N${_\text{tot}}$) after 500 fs for each \textup{\textsmaller{L}}-lysine (Lys) crystal system (orange plot). Experimental (Exp.) external quantum efficiencies (Ext. Quant. Eff.) of Lys acidified with HCl (green) and H$_2$SO$_4$ (purple) are shown for comparison. Panel b shows the calculated lowest absorption energies (Exc., vertical bars) and emission spectra (Emis., solid curves) for LLS (blue/red), LLMHCl (magenta/green), and LLH (yellow/cyan). Vertical dashed lines indicate the emission maxima. Panel c depicts the molecular orbitals involved in the lowest-energy excitation of LLS: HOMO (left) and LUMO (right), with positive and negative orbital lobes colored in green and purple, respectively.}
    \label{fig:Fig6}
\end{figure}

Fig. \ref{fig:Fig6}b presents the calculated fluorescence spectra of LLS (red line), LLMHCl (green line), and LLH (cyan line) averaged over the trajectories remaining in S$_1$. Vertical lines indicate the averaged lowest absorption (excitation) transitions: blue (LLS), magenta (LLMHCl), and yellow (LLH). For all systems, excitation wavelengths cluster near $\sim210$ nm, exhibiting a $\sim$120 nm ($\sim 2.1$ eV) blue-shift relative to experiment. This discrepancy likely arises from a combination of known TDDFT limitations, such as the tendency of CAM-B3LYP to overestimate vertical excitation energies\cite{Shao_2020}, as well as neglect of nuclear quantum effects (NQEs), and structural defects or disorder\cite{Meirzadeh_2018,Kurashvili_2022}. Several previous studies have shown that the inclusion of NQEs can alter both the peak absorption in organic systems by up to $\sim$1 eV, as well as affect the line-shape\cite{Feher_2021,Sappati_2016,Pinotsi2016,Law_2017}. In addition, the size of the QM subsystem can also influence spectral positions. While our simulations employed QM dimers as a compromise between accuracy and computational cost, benchmarks in other amino acid crystals\cite{Banerjee_2025} have shown that extending to larger clusters, such as QM heptamers, red-shifts absorption energies by $\sim0.5$ eV, thereby reducing the discrepancy with experiment.

Across protonation states, the $\text{S}_0 \rightarrow\text{S}_1$ transition consistently occurs from the highest occupied molecular orbital (HOMO) to the lowest unoccupied molecular orbital (LUMO), which exhibit n and $\pi^*$ character, respectively. In LLS (Fig. 6c), the n-type HOMO localizes on the C-terminal carboxyl and partly on SO$_4^{2-}$ oxygens, while the $\pi^*$-type LUMO resides on the same Lys monomer, primarily on the C-terminal oxygen and along the CC$_\alpha$ bond. Similar localizations occur for LLMHCl (Fig. \ref{fig:SI_7}a) and LLH (Fig. \ref{fig:SI_8}a), with the n-type HOMO on CO$_2^{-}$ oxygens and the $\pi^*$-type LUMO on the same group and adjacent C$_\alpha$. Cl$^-$ ions show no active role in LLMHCl. In the LLH crystal, both frontier orbitals are confined to the C-terminus engaged in head-to-head HBs, with no contribution from the –NH$_2$ side chain group within our QM description. Therefore, our results indicate that the dominant photoexcited electronic structure arises from the C-terminal hydrogen bonding motif. Finally, in all systems, the electron density alternates between the two QM Lys monomers, consistent with dimer symmetry, and localizes within H-bonding regions, linking local non-covalent interactions with photoexcited electronic properties. Comparison with the isolated monomer (shown for LLH as an example in Fig. \ref{fig:SI_9}) highlights that the solid-state packing and HB network substantially influence the nature of the excitation. Removing the crystal environment converts the transition to an n $\rightarrow \sigma^*$ character, with HOMO and LUMO localized on the C- and N-terminus, respectively, of the same Lys monomer, illustrating the role of aggregation in shaping the electronic structure.

While excitation energies are relatively pH-independent, emission spectra show more pronounced differences. The computed emission trend in Fig. \ref{fig:Fig6}b mirrors the one observed experimentally for Lys samples acidified with H$_2$SO$_4$ (Fig. \ref{fig:Fig4}b). LLS (pH 1) exhibits the most blue-shifted emission ($\lambda{\text{max}} = 320$ nm), followed by LLMHCl (pH 7) and LLH (pH 10), both exhibiting emission maxima near 356 nm. LLMHCl displays an additional broad 460-480 nm feature, originating from structures in which ES distortions--particularly C$\alpha$CO$_2$ deplanarization--are partially activated but insufficient to reach the conical intersection (CoIn). This mode, typically associated with non-radiative decay\cite{miron2023}, in this case gives rise to red-shifted emission from distorted yet emissive geometries. LLS emission instead exhibits a shoulder near the excitation energy, suggesting an ES minimum near the Franck-Condon region. Although emission energies remain overall blue-shifted by 50–80 nm (0.5-0.7 eV) relative to experiment, this discrepancy is significantly reduced compared to excitation, and spectral tails at longer wavelengths ($\lambda>550$ nm) are well captured for LLMHCl and LLH. All S$_1\rightarrow$S$_0$ transitions retain the same n–$\pi^*$ character as the GS excitations (Figs. \ref{fig:SI_7}, \ref{fig:SI_8}, and \ref{fig:SI_10}). 

To facilitate comparison with the normalized experimental emission spectra (see above), the calculated emission spectra in Fig. 6b were also normalized. However, normalization prevents direct comparison of absolute intensities. Therefore, we examined the distributions of oscillator strength (f) values, which measure the radiative probability of electronic transitions, also complementing the previous S$_1$ population analysis (Fig. 6a). Normalized histograms and Gaussian fits for LLS, LLMHCl, and LLH are shown in Fig. \ref{fig:SI_11}a-c, with an overlay of the fitted curves in Fig. \ref{fig:SI_11}d. For LLH (cyan) and LLMHCl (green), the distributions peak below 0.001, with minor tails up to 0.005. Although these values are modest--well below those typical of aromatic fluorophores\cite{GDM_2024}--their overall contribution to the emission is amplified by the large fraction of trajectories remaining in S$_1$ (Fig. 6a). In contrast, LLS (red) shows a broader distribution extending to $\sim$0.01, but its emission remains minimal because only a small fraction of trajectories survive in S$_1$ long enough to emit, due to dominant non-radiative decay pathways.

\subsubsection{Vibrational Modes Initiating Non-Radiative Decay}

To understand why the LLS crystal exhibits particularly weak fluorescence, we analyzed the vibrational modes that funnel the system to the CoIn and, subsequently, enable decay to the ground state. Previous studies\cite{Pinotsi2016,Sijo_2019,Grisanti2020,Stephens2021,miron2023} have identified proton transfer, carbonyl stretching, and deplanarization of the amide group as key modes governing the ES lifetimes of peptide chains, amyloid-like peptides, and glutamine crystals. Constraining these modes raises the energy barrier to the CoIn, thereby prolonging the ES lifetime. More recently, Banerjee and co-workers\cite{Banerjee_2025} linked the weaker fluorescence of cysteine crystals grown in H$_2$O, as opposed to D$_2$O, to activation of C$_\alpha$N and SH stretching modes.

\begin{figure}[h!]
    \centering
    \includegraphics[width=\linewidth]{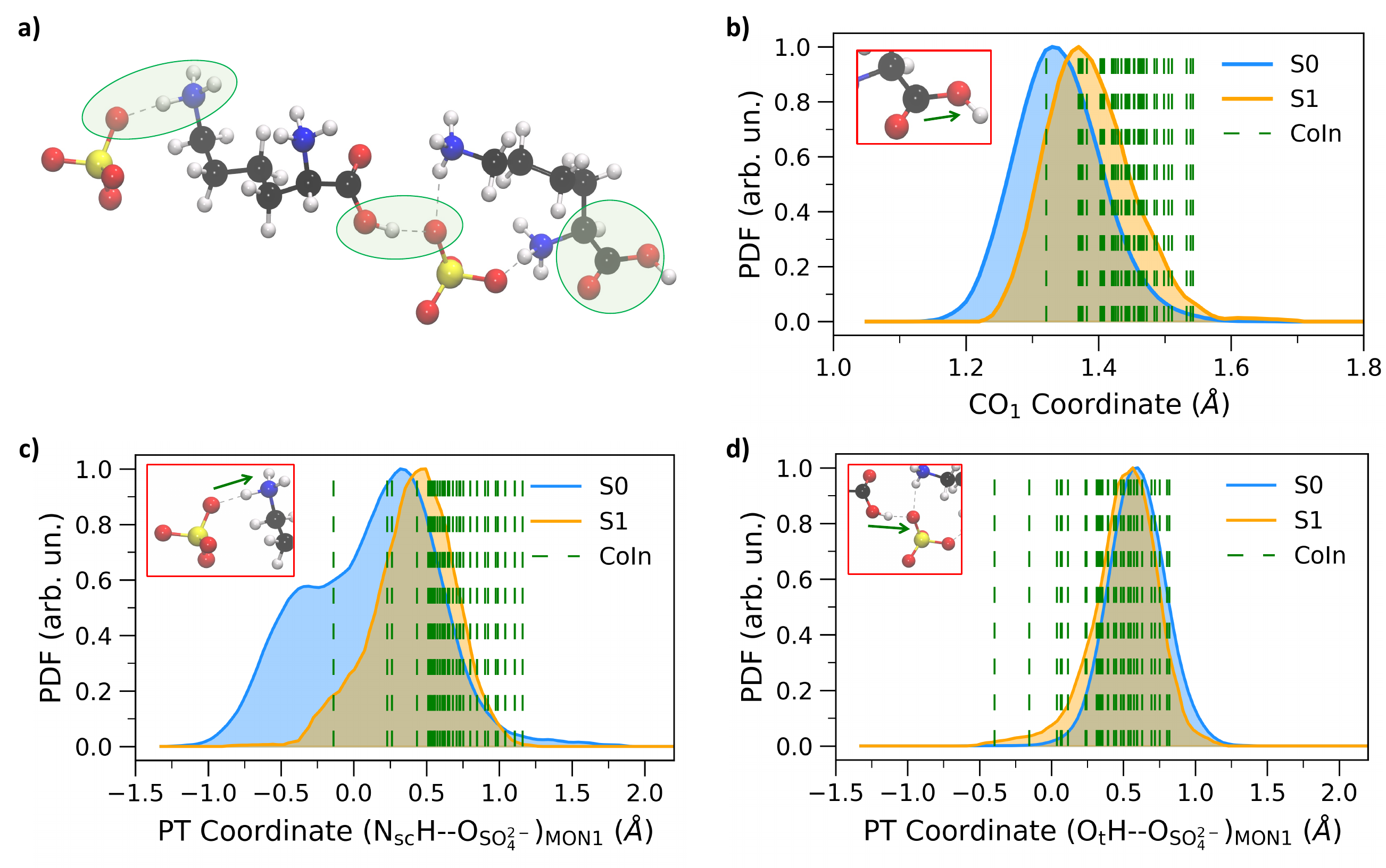}
    \caption{Probability distribution functions (PDFs) for the vibrational coordinates that govern non-radiative decay in (36) surface-hopping trajectories of \textup{\textsmaller{L}}-lysine sulfate (LLS). Panel (a): CPK representation of the quantum-mechanical (QM) region used in the simulations, consisting of a Lys dimer and two sulfate ions. Green circles and ellipses highlight the atomic regions involved in key vibrational modes. PDFs are reported for: (b) C=O stretching mode; (c-d) proton transfer (PT) coordinates, defined as d$_{\text{X}\text{H}}-$d$_{\text{O}\text{H}}$, where X can be N$_{\text{sc}}$ (c), or O$_\text{t}$ (d), while O corresponds to the oxygen atom of the sulfate ion. PDFs corresponding to the ground state S$_0$ are plotted in blue; those for the first excited state S$_1$ are plotted in orange. Dashed green vertical lines indicate the PT coordinate values at which the conical intersection (CoIn) is reached, and non-radiative decay occurs from S$_1\rightarrow$ S$_0$. Regions of the LLS dimer involved in the modes are shown in the insets. In the PDF plots of the different PT coordinates, MON1 and MON2 refer to the QM Lys monomer on the left and right (panel a), respectively.}
    \label{fig:Fig7}
\end{figure}

Guided by the n $\rightarrow \pi^*$ character of LLS (Fig. \ref{fig:Fig6}c), we first focused on vibrational modes involving the C-terminus (Figs. \ref{fig:SI_12}-\ref{fig:SI_13} for the two QM Lys monomers). We then analyzed coordinates associated with the system hydrogen bonding network (Fig. \ref{fig:SI_14}). From this combined set, a few modes emerge as the dominant funnels to the CoIn, as shown in Fig. \ref{fig:Fig7}. 

The collective motion driving non-radiative decay in LLS (Fig. \ref{fig:Fig7}a) can be decomposed into two main contributions: (i) stretching of the CO bond coupled with deplanarization of the C$_\alpha$CO$_2$ group, and (ii) PT coordinates involving NH or OH groups of Lys and sulfate oxygens. As shown in Fig. \ref{fig:Fig7}b, the CO bond elongates in S$_1$ (orange distribution), increases further at the CoIn (dashed line), and contracts upon relaxation to S$_0$ (blue distribution). A similar trend is observed for the second CO bond of the carboxylate group (Fig. \ref{fig:SI_13}c). This distortion is concerted with amide deplanarization in S$_1$, followed by replanarization after decay (Fig. \ref{fig:SI_13}d). These features are consistent with previously reported ES dynamics in non-emissive amino acid crystals, such as \textup{\textsmaller{L}}-glutamine\cite{miron2023}.

However, unlike glutamine--where CO stretching is triggered by transient strengthening (PT coordinate $<0$) of a -NH$_3^+\cdots$CO HB between monomers--LLS exhibits HBs exclusively with sulfate counterions. Consequently, CO bond activation and proton transfer may occur in different structural units. The primary HB-related mode promoting non-radiative decay involves transient weakening of an -NH$_3^+\cdots$SO$_4^{2-}$ HB in S$_1$, particularly near the CoIn, with PT typically occurring upon relaxation to S$_0$ (Figs. \ref{fig:Fig7}c and \ref{fig:SI_14}f). A secondary PT mode involves the C-terminal OH group and displays a slight transient HB strengthening in S$_1$, with occasional proton sharing or transfer to the sulfate oxygen at the CoIn, and reversion after decay (Fig. \ref{fig:Fig7}d). Many trajectories that reach the CoIn show PT coordinates below $0.35$ $\angstrom$, reflecting a weakened interaction. Collectively, these results highlight that CO bond distortions and HB fluctuations actively govern the accessibility of the CoIn under acidic conditions, with the sulfate anion also playing a crucial role in this dynamic behavior.

Similar CO elongation and C$_\alpha$CO$_2$ deplanarization appear in the rare non-radiative trajectories of LLMHCl and LLH (Figs. \ref{fig:SI_15}–\ref{fig:SI_16}b–d), but no PT occurs in S$1$ or S$0$. In both systems, weakening of the NH$\dotsm$O coordinates is observed, with values often exceeding $1$ $\angstrom$, but stems from other structural distortions driving the non-radiative decay. In LLMHCl, CO stretching and deplanarization lead to C$_\alpha$C elongation in S$0$ (Fig. \ref{fig:SI_15}e), increasing the NH–acceptor distance (Fig. \ref{fig:SI_15}f). In LLH, significant C$_\alpha$N stretching near the CoIn weakens one HB, while the other is disrupted by flipping of the carboxylate group (Figs. \ref{fig:SI_16}f–h). This C$_\alpha$N mode was recently implicated in the weak fluorescence of cysteine crystals.\cite{Banerjee_2025} 

These decay events, though rare, indicate that under neutral and basic conditions, non-radiative decay is primarily driven by backbone distortions rather than HB dynamics. In contrast, under acidic conditions, the combination of enhanced protonation and the presence of flexible, multiply charged counterions--such as sulfate ions--increases the dynamical complexity of the hydrogen bonding network, opening new non-radiative pathways.

\section{Conclusions}

In conclusion, we combined experiments and theory to investigate the evolution of fluorescence in solid-state lysine as a function of pH. To the best of our knowledge, the physical factors and mechanisms controlling intrinsic fluorescence in non-aromatic supramolecular assemblies under varying pH remain poorly understood and have been explored in only a few studies\cite{Pinotsi2016,stagi2022,Wang_2025}. One of our key findings is that both pH and the counterions introduced by the acids used to adjust it modulate protonation states and hydrogen bonding networks, altering the physical and chemical properties of the assemblies.

Using a synergy of microscopy techniques, we reveal subtle yet distinct morphological changes in lysine aggregates from acidic to basic conditions, which correlate with enhanced fluorescence intensity at higher pH. To rationalize these trends, we performed excited state molecular dynamics simulations on lysine crystal models, revealing that non-radiative decay pathways are more accessible at lower pH, consistent with experiments. Specifically, under acidic conditions, the presence of a counterion forming strong hydrogen bonds with the amino-acid, activates channels--including proton transfer--that facilitate easy access to the conical intersection.

Our findings open up important perspectives and challenges on both experimental and theoretical fronts. Specifically, advancing experimental methods to better characterize local and macroscopic structures of titratable amino acid aggregates as a function of pH remains a crucial factor to guide computational modeling. While our models capture qualitative trends, a unique aspect we uncovered in this study is that the non-radiative decay paths under acidic conditions involve proton transfer processes for both normal and strong hydrogen bonds. An important implication of this is that the choice of the QM/MM boundary may suppress or enhance proton transfer events. While we have found that our conclusions are robust to this factor (see Fig. \ref{fig:SI_17} in the SI), efforts that couple reactive QM/MM boundaries for excited state dynamics will be interesting to pursue in the future. Moreover, in all the investigated crystals, we have limited the QM region to Lys dimers, which sufficed to capture the mechanistic trends, in line with previous studies\cite{Banerjee_2025}. Future computational work using larger QM regions or including reactive water molecules could provide improved quantitative agreement with experiments, while further elucidating the atomistic origins of intrinsic fluorescence in non-aromatic supramolecular assemblies.



\begin{acknowledgement}

MM, GDM, DB, and AH acknowledge the European Commission for funding on the ERC Grant HyBOP 101043272. They also thank CINECA supercomputing (project NAFAA-HP10B4ZBB2) and MareNostrum5 (project EHPC-EXT-2023E01-029) for computational resources. CD, AA, and LV acknowledge the project PRIN$\_$2022TSLMHR entitled “Biomaterials from peptide self-assembling generated by mimicking protein amyloid-like structures”, funded by Ministero dell’Universit$\text{à}$ e della Ricerca - NextGenerationEU (European Union). LC acknowledges Ricerca Corrente Project.

\end{acknowledgement}

\section*{Conflict of Interest}
The authors declare no conflict of interest.

\clearpage

\begin{suppinfo}
\beginsupplement
Critical aggregation concentration measurements; birefringence microscopy images of Lys aggregates acidified with HCl and H$_2$SO$_4$ at pH 10; confocal fluorescence images of Lys aggregates prepared using HCl; normalized steady-state fluorescence spectra of Lys aggregates under varying excitation wavelengths (HCl and H$_2$SO$_4$); excitation-emission correlation plots; molecular orbitals involved in excitation and emission of \textup{\textsmaller{L}}-Lysine sulfate (LLS), \textup{\textsmaller{L}}-Lysine monohydrochloride dihydrate (LLMHCl), and \textup{\textsmaller{L}}-Lysine hemihydrate (LLH); oscillator strength distributions for the three Lys crystal forms; vibrational coordinate analyses from trajectories undergoing non-radiative decay (LLS, LLMHCl, LLH); validation of the QM/MM setup for LLS; and additional references.

\section{Supplementary Text}

\subsection{Critical Aggregation Concentration}
\subsubsection{Experimental Details}
The intrinsic fluorescence of lysine occurs only in the aggregated state while is absent at dilute concentrations\cite{stagi2022}. Therefore, it is essential to verify that experimental conditions are above the critical aggregation concentration (CAC). In this work, the CAC value of Lys in water was determined by using fluorescence spectroscopy with two established probe molecules: (a) 8-anilino-1-naphthalene sulfonic acid ammonium salt (ANS) and (b) pyrene (pyr). Fluorescence spectra were recorded at room temperature on a Jasco spectrofluorophotometer (Model FP-8550) using a quartz cuvette (1.0 cm path length). Experimental parameters were: excitation and emission bandwidths = 5 nm; recording speed = 120 nm$\cdot$min$^{-1}$, excitation wavelength = 350 nm (ANS) and 335 nm (pyr). For titrations, increasing aliquots of Lys solution were added to 200 $\mu$L of 200 $\mu$M ANS or 10 $\mu$M pyr aqueous solutions. Blank spectra were subtracted, and fluorescence intensities were corrected for dilution. Titrations were performed in triplicate, and trends were analyzed by least-squares fitting. In the ANS assay, the CAC value was determined from a linear least-squares fit of the fluorescence intensity at 475 nm (excitation at 350 nm) versus Lys concentration, following Balasco \textit{et al.}\cite{Balasco2024}. In the pyr assay, the CAC was determined from the break-point of the fitting of the I1/I3 ration (excitation at 335 nm) versus Lys concentration\cite{MANJU2011318}.

\subsubsection{CAC Measuraments}
\textup{\textsmaller{L}}-Lysine (Lys) samples for the experimental morphological and optical characterization were prepared by dissolving the amino acid powder in double distilled water at the concentration of 1 g/mL (6.8 mol/L), which corresponds to the maximum of its solubility. Under these conditions, an increase of the volume of $\sim70\%$ was observed, and the resulting solution exhibited a slight yellow coloration. The pH of the resulting solution was basic (pH = 11.8). Given the known ability of single amino acids to self-assemble into supramolecular structure\cite{GOUR2021154}, the formation of Lys aggregates was investigated using two environment-sensitive fluorescent probes: ANS and pyr, as shown in Fig. \ref{fig:SI_2}\cite{diaferia_2015}. Both dyes are known to be able to change their fluorescence properties as function of the surrounding chemical environment (hydrophobic or hydrophilic). ANS shows a characteristic emission peak between 460-480 nm in hydrophobic environments, such as the inner micellar compartment\cite{Vaccaro_2007} or aromatic peptide aggregate spine\cite{Diaferia_2016}. The fluorescence of pyr, on the other hand, is characterized by five vibrational bands, where the intensity ratio between the first (I1 at 373 nm) and third (I3 at 383 nm) bands serves as a sensitive indicator of environmental polarity. 

The CAC of Lys was estimated from the inflection points of fluorescence titration curves (Fig. \ref{fig:SI_2}). CAC values determined using ANS (0.915 mol/L) and pyr (0.888 mol/L) fluorophores are in good agreement, supporting the reliability of the measurements. These values are slightly higher than the previously reported estimate of 0.5 mol/L by Stagi \textit{et al.} \cite{stagi2022}, who identified a change in the slope of the lysine fluorescence intensity as a function of its concentration: at lower concentrations, intensity increased linearly, while at higher concentrations (from 0.4 mol/L), the rate of increase slowed, suggesting the formation of larger, less fluorescent aggregates. The fact that the ANS and pyr calculated values are below the Lys solubility (solubity/CAC $\sim$7.5) confirms that Lys aggregates, or smaller clusters, are present under the experimental conditions investigated. 

The CAC was determined only under basic conditions (pH = 11.8), corresponding to the zwitterionic form of Lys, to confirm that the amino acid is aggregated under the experimental conditions used in this work. However, since pH strongly influences the structural and emissive properties of Lys, a more systematic investigation of the CAC as a function of pH would be an interesting direction for future studies.

\clearpage
\section{Supplementary Figures}
    \begin{figure}[H]
    \centering
    \includegraphics[width=\linewidth]{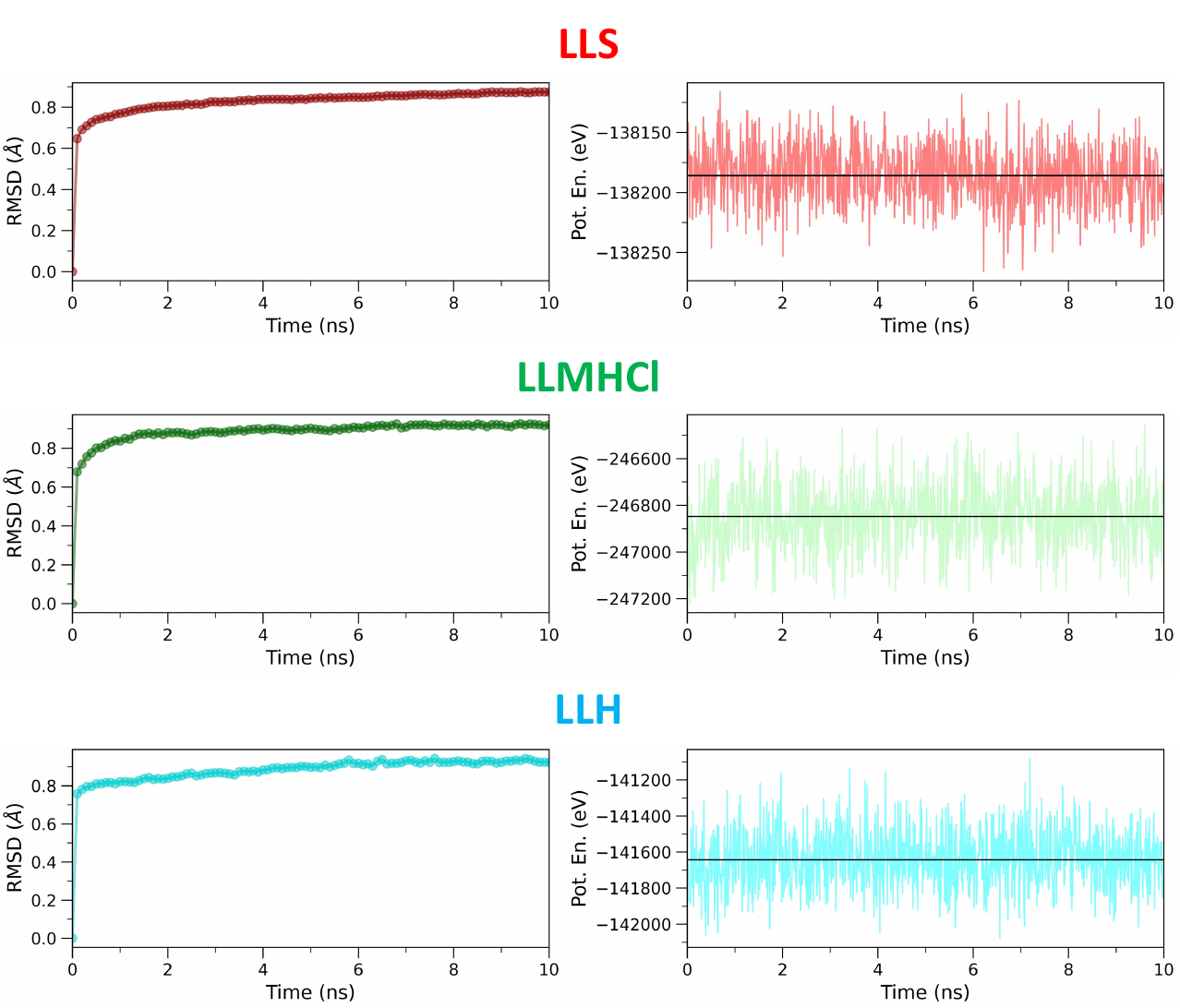}
    \caption{Root mean square deviation (RMSD, left) and potential energy (Pot. En., right) along the 10 ns NVT equilibration of the Lys crystal supercells. Top panels: \textup{\textsmaller{L}}-Lysine sulfate (LLS, red); middle panels: \textup{\textsmaller{L}}-Lysine monohydrochloride (LLMHCl, green); bottom panels: \textup{\textsmaller{L}}-Lysine hemihydrate (LLH, cyan). RMSD is calculated relative to the first frame of each trajectory.}
    \label{fig:SI_1}
    \end{figure}

\begin{figure}[H]
    \centering
    \includegraphics[width=0.7\textwidth]{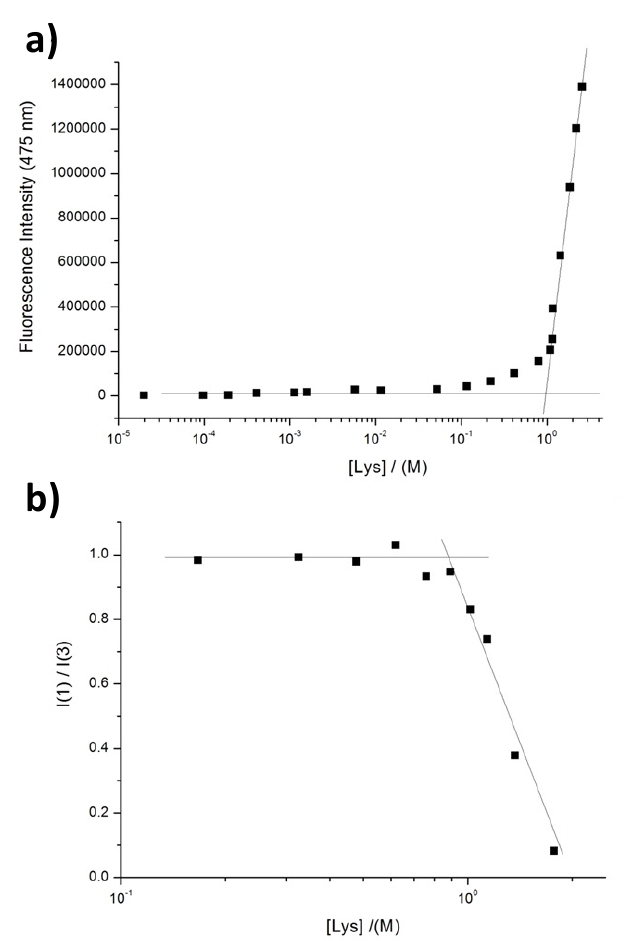}
    \caption{Fluorescence titration of 200 $\mu$M ANS (A) and 10 $\mu$M pyr (B) with increasing amount of Lys. Critical aggregation concentration values are established from the graphical break points of the titrations.}
    \label{fig:SI_2}
\end{figure}

\begin{figure}[H]
    \centering
    \includegraphics[width=\linewidth]{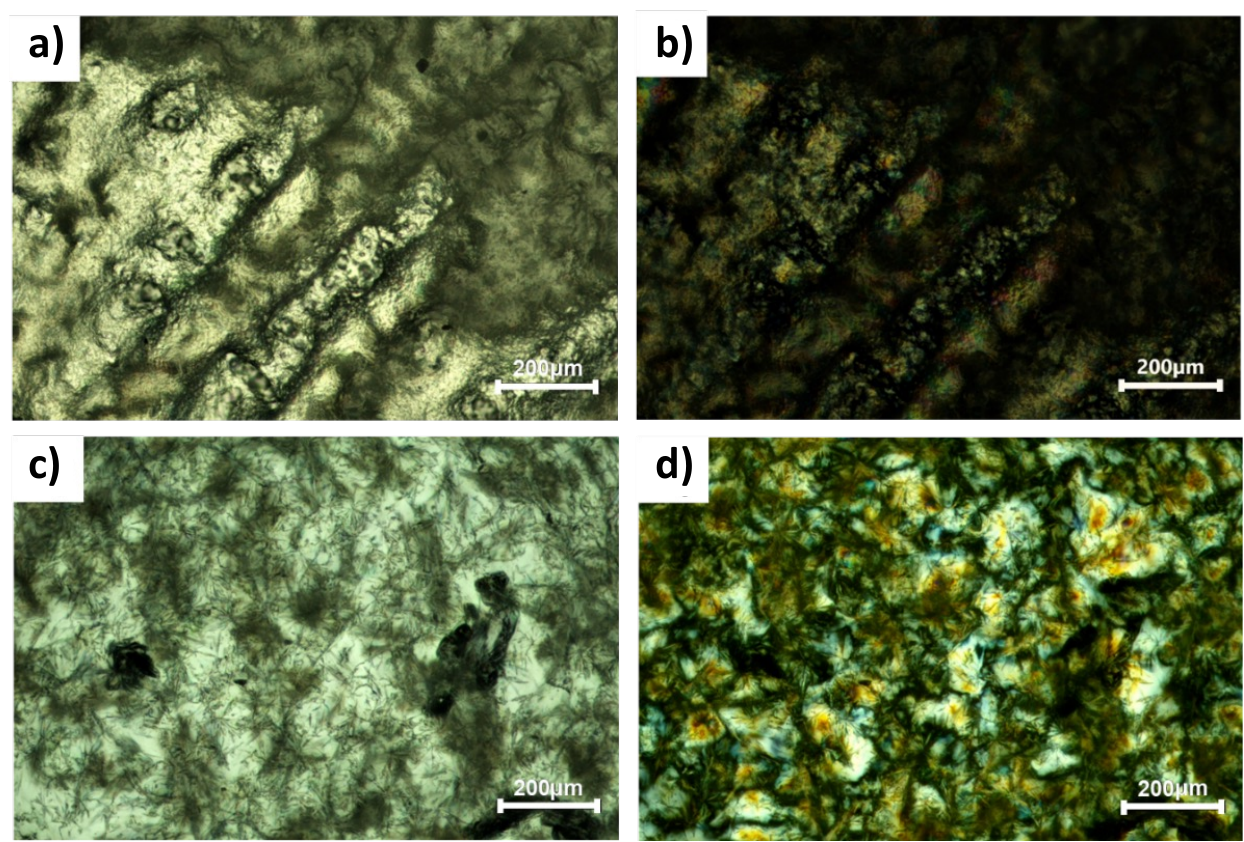}
    \caption{Panel a and b show the birefringence of Lys acidified with HCl at pH = 10 under bright field (left, panel a) and cross-polarized light (right, panel b). Similar results are reported for Lys acidified with H$_2$SO$_4$ in panel c and d.}
    \label{fig:SI_3}
\end{figure}

\begin{figure}[H]
    \centering
    \includegraphics[width=\linewidth]{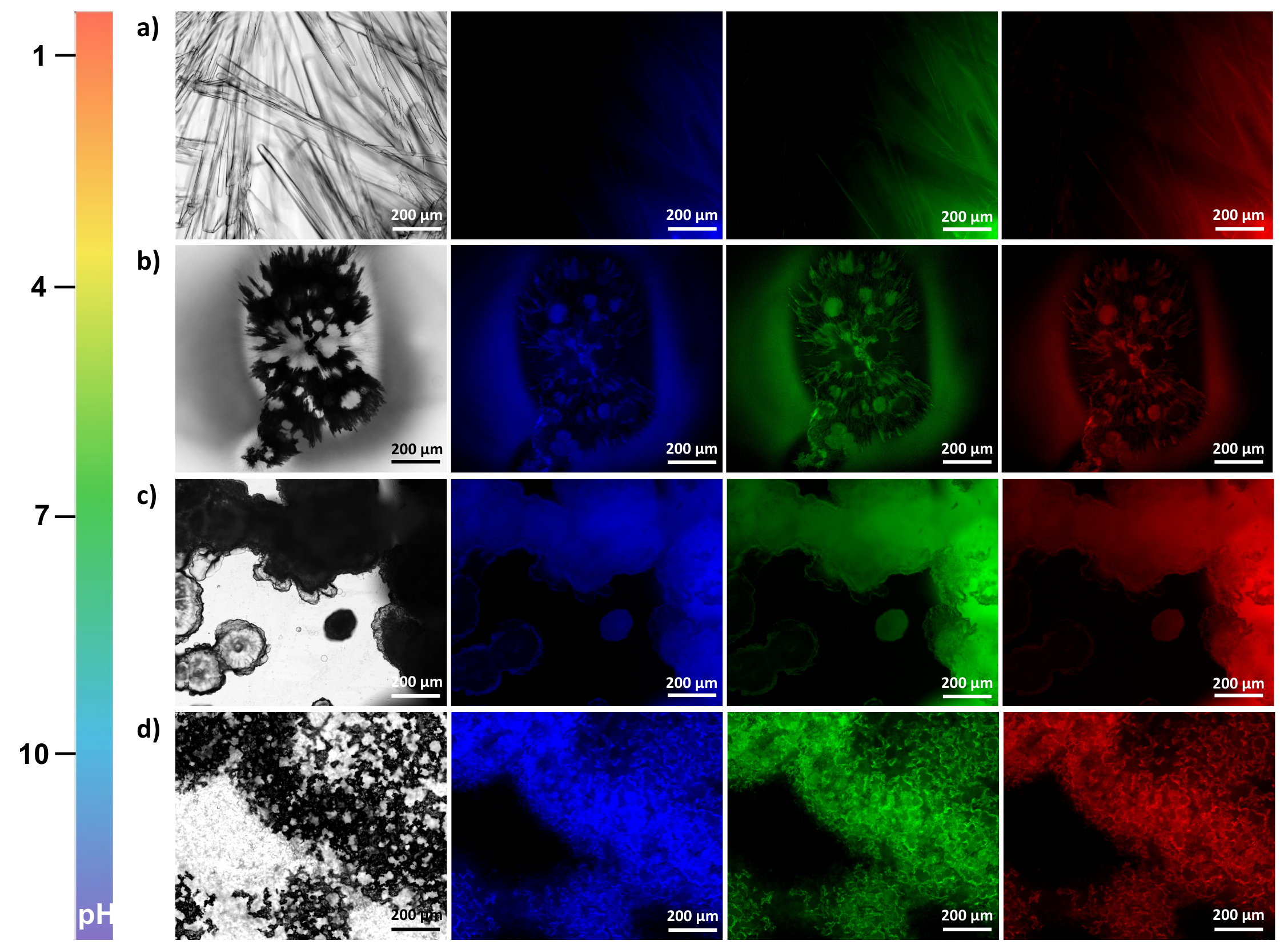}
    \caption{Confocal microscopy images of Lys aggregates formed at pH 1, 4, 7, and 10 using HCl as the acidifying agent are shown in panel a, b, c, and d, respectively. For each pH, images are arranged from left to right as follows: panel (a) begins with an optical microscopy image (used in place of dark-field due to unavailability), followed by fluorescence images recorded in the blue ($\lambda_{\text{exc}}$ = 402 nm, $\lambda_{\text{emi}}$ = 421 nm), green ($\lambda_{\text{exc}}$ = 495 nm, $\lambda_{\text{emi}}$ = 519 nm), and red ($\lambda_{\text{exc}}$ = 590 nm, $\lambda_{\text{emi}}$ = 617 nm) channels. Panels (b–d) begin with dark-field images followed by the same fluorescence channels.}
    \label{fig:SI_4}
\end{figure}

\begin{figure}[H]
    \centering
    \includegraphics[width=0.7\linewidth]{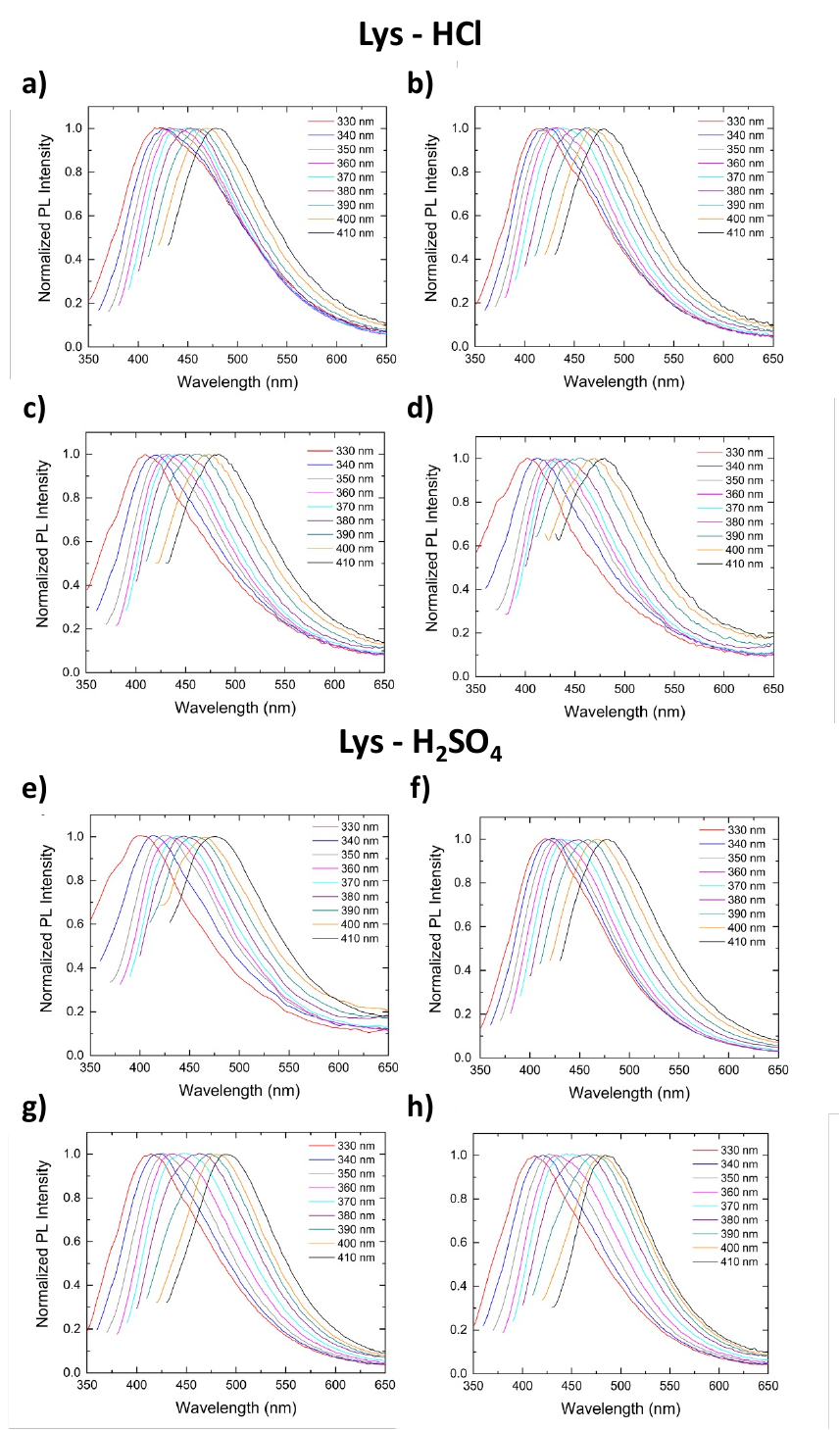}
    \caption{Normalized fluorescence spectra of Lys in HCl (a-d) and H$_2$SO$_4$ (e-h) at varying pH values, plotted as a function of excitation wavelength (330–410 nm). Spectra were normalized to their maximum emission intensity to highlight shifts in peak positions with increasing excitation wavelength. Panels correspond to different pH values: (a, e) pH 1; (b, f) pH 4; (c, g) pH 7; (d, h) pH 10.}
    \label{fig:SI_5}
\end{figure}

\begin{figure}[H]
    \centering
    \includegraphics[width=\linewidth]{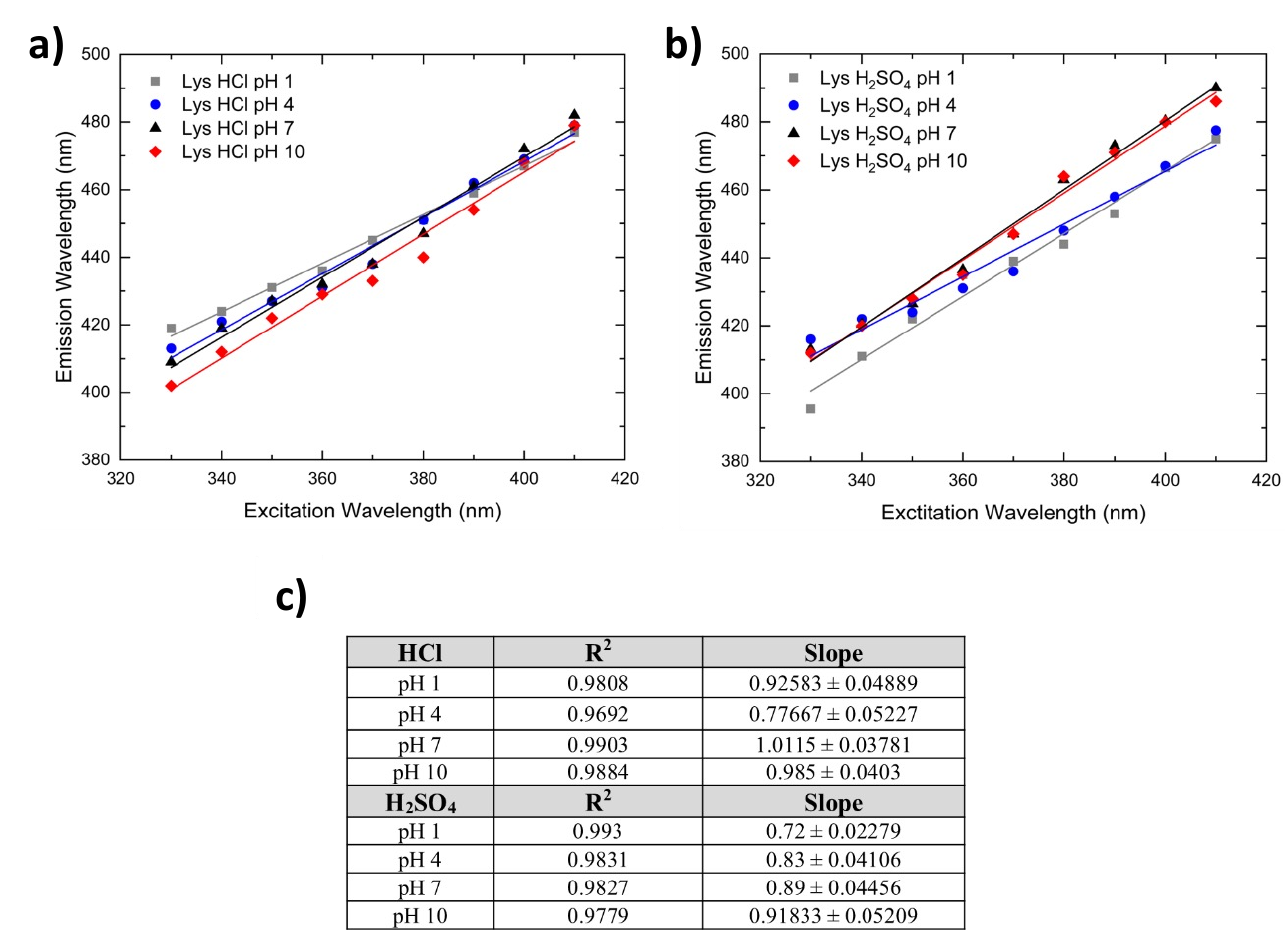}
    \caption{Plots of the maxima positions as function of the excitation wavelength at different pH conditions (1, 4, 7, and 10) for Lys acidified with HCl (panel a) and H$_2$SO$_4$ (panel b). R$^2$ and slope values of the linear best-fit curves are reported in the table in panel c.}
    \label{fig:SI_6}
\end{figure}
\clearpage

\begin{figure}[H]
    \centering
    \includegraphics[width=\linewidth]{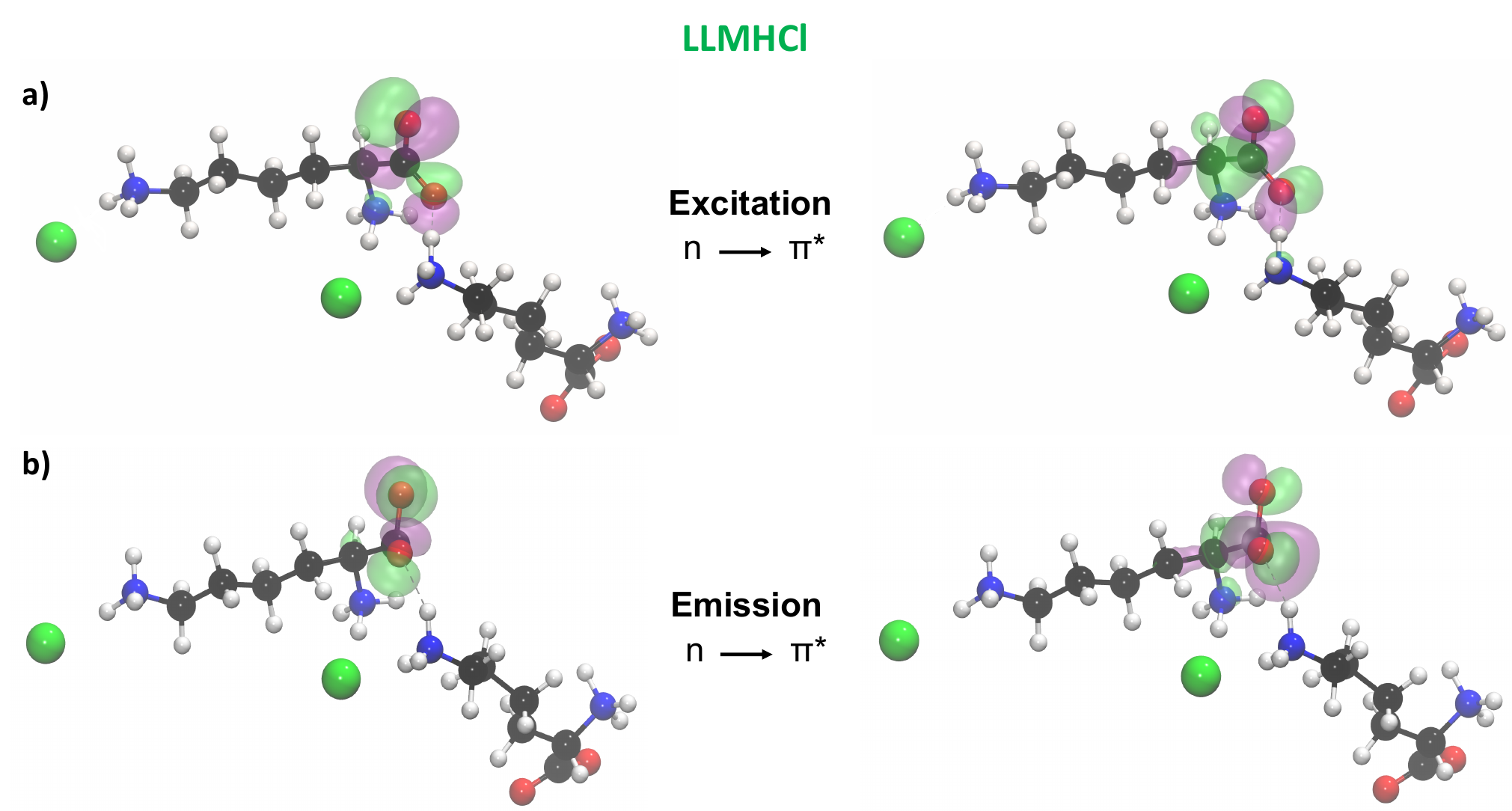}
    \caption{Molecular orbitals involved in the (a) excitation (S$_0 \rightarrow$ S$_1$) and (b) emission (S$_1 \rightarrow$ S$_0$) transitions of \textup{\textsmaller{L}}-Lysine monohydrochloride dihydrate (LLMHCl). Both transitions show an n $\rightarrow$ $\pi^*$ character. The highest occupied molecular orbital (HOMO) and lowest unoccupied molecular orbital (LUMO) are shown on the left and right, respectively. For consistency, the emission is depicted using the same orbital configuration as the excitation (i.e., in the S$_0 \rightarrow$ S$_1$ direction). Positive and negative lobes of the MOS are shown in green and purple, respectively.}
    \label{fig:SI_7}
\end{figure}

\clearpage
\begin{figure}[H]
    \centering
    \includegraphics[width=\textwidth]{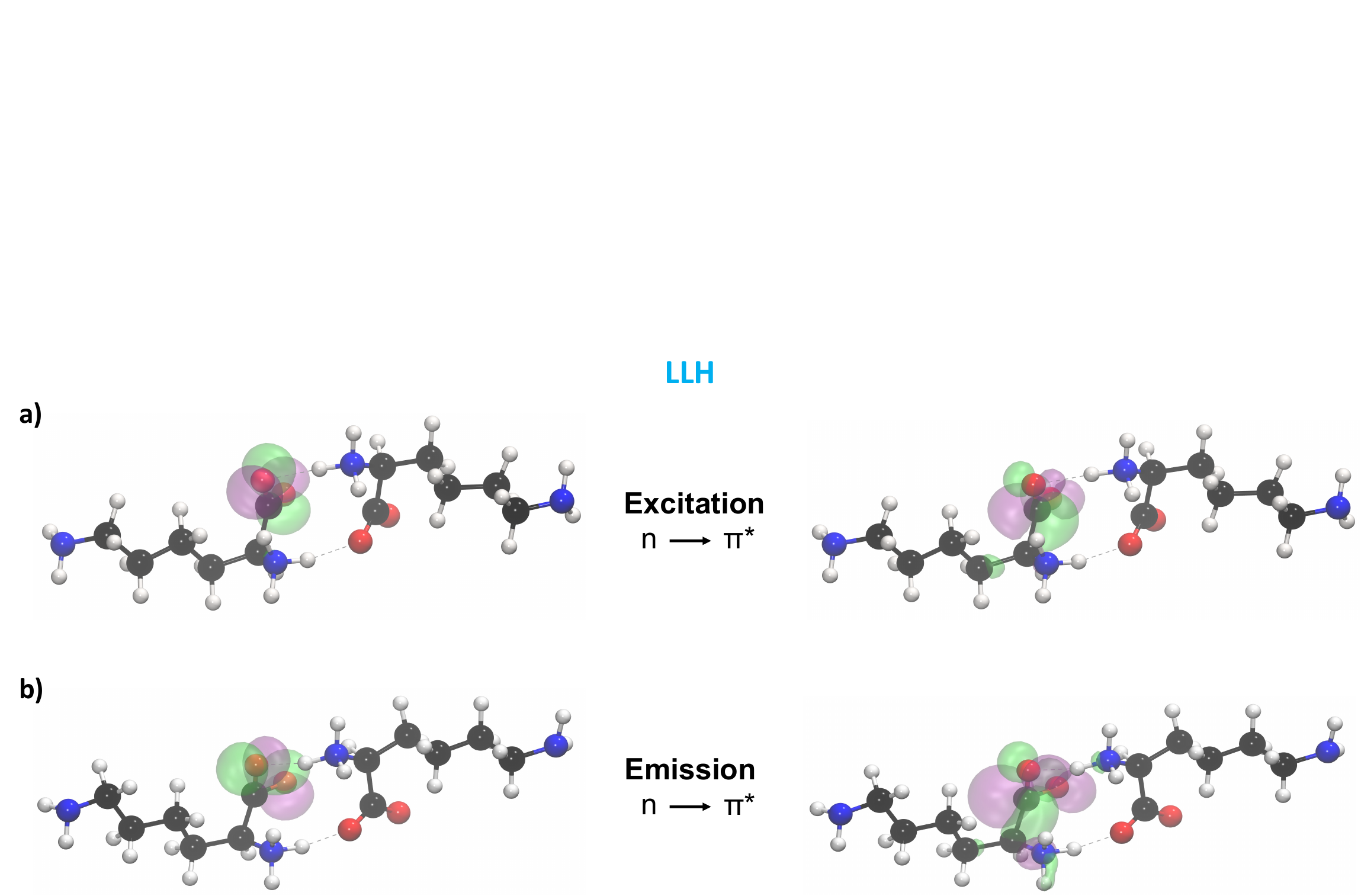}
    \caption{Molecular orbitals involved in the (a) excitation (S$_0 \rightarrow$ S$_1$) and (b) emission (S$_1 \rightarrow$ S$_0$) transitions of \textup{\textsmaller{L}}-Lysine hemihydrate (LLH). Both transitions show an n $\rightarrow$ $\pi^*$ character. The highest occupied molecular orbital (HOMO) and lowest unoccupied molecular orbital (LUMO) are shown on the left and right, respectively. For consistency, the emission is depicted using the same orbital configuration as the excitation (i.e., in the S$_0 \rightarrow$ S$_1$ direction). Positive and negative lobes of the MOS are shown in green and purple, respectively.}
    \label{fig:SI_8}
\end{figure}

\clearpage

\begin{figure}[H]
    \centering
    \includegraphics[width=\linewidth]{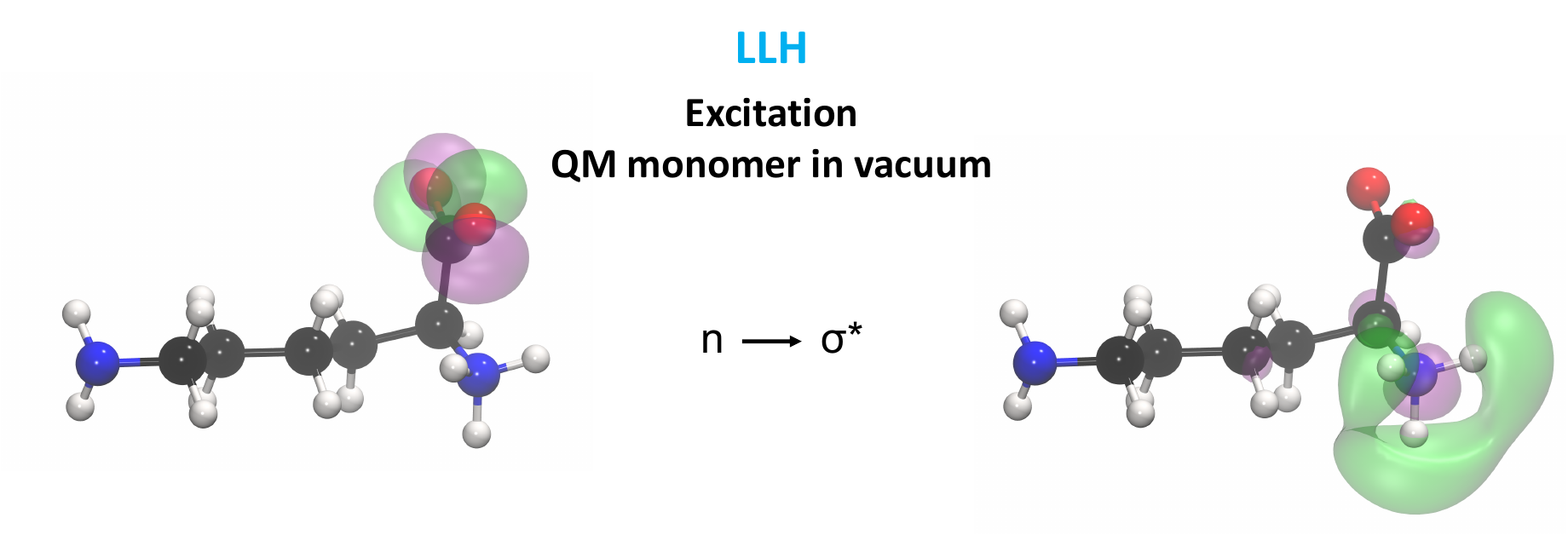}
    \caption{Molecular orbitals involved in the excitation (S$_0 \rightarrow$ S$_1$) transition of a single \textup{\textsmaller{L}}-Lysine hemihydrate (LLH) monomer in vacuum. The transition shows an n $\rightarrow$ $\sigma^*$ character. The highest occupied molecular orbital (HOMO) and lowest unoccupied molecular orbital (LUMO) are shown on the left and right, respectively, with positive and negative lobes depicted in green and purple.}
    \label{fig:SI_9}
\end{figure}
\clearpage

\begin{figure}[H]
    \centering
    \includegraphics[width=\linewidth]{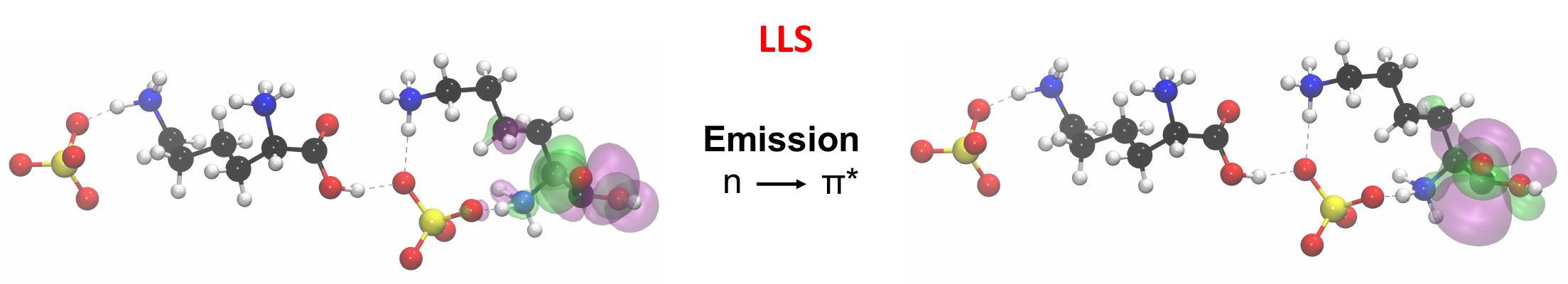}
    \caption{Molecular orbitals (MOs) involved in the  emission (S$_1 \rightarrow$ S$_0$) transition of \textup{\textsmaller{L}}-Lysine sulfate (LLS). For consistency, the emission is depicted using the same orbital configuration as the excitation (i.e., in the S$_0 \rightarrow$ S$_1$ direction), which is shown in Fig. 4f in the main text. Both transitions show an n $\rightarrow$ $\pi^*$ character. The highest occupied molecular orbital (HOMO) and lowest unoccupied molecular orbital (LUMO) are shown on the left and right, respectively. Positive and negative lobes of the MOS are shown in green and purple, respectively.}
    \label{fig:SI_10}
\end{figure}
\clearpage

\begin{figure}[H]
    \centering
    \includegraphics[width=\linewidth]{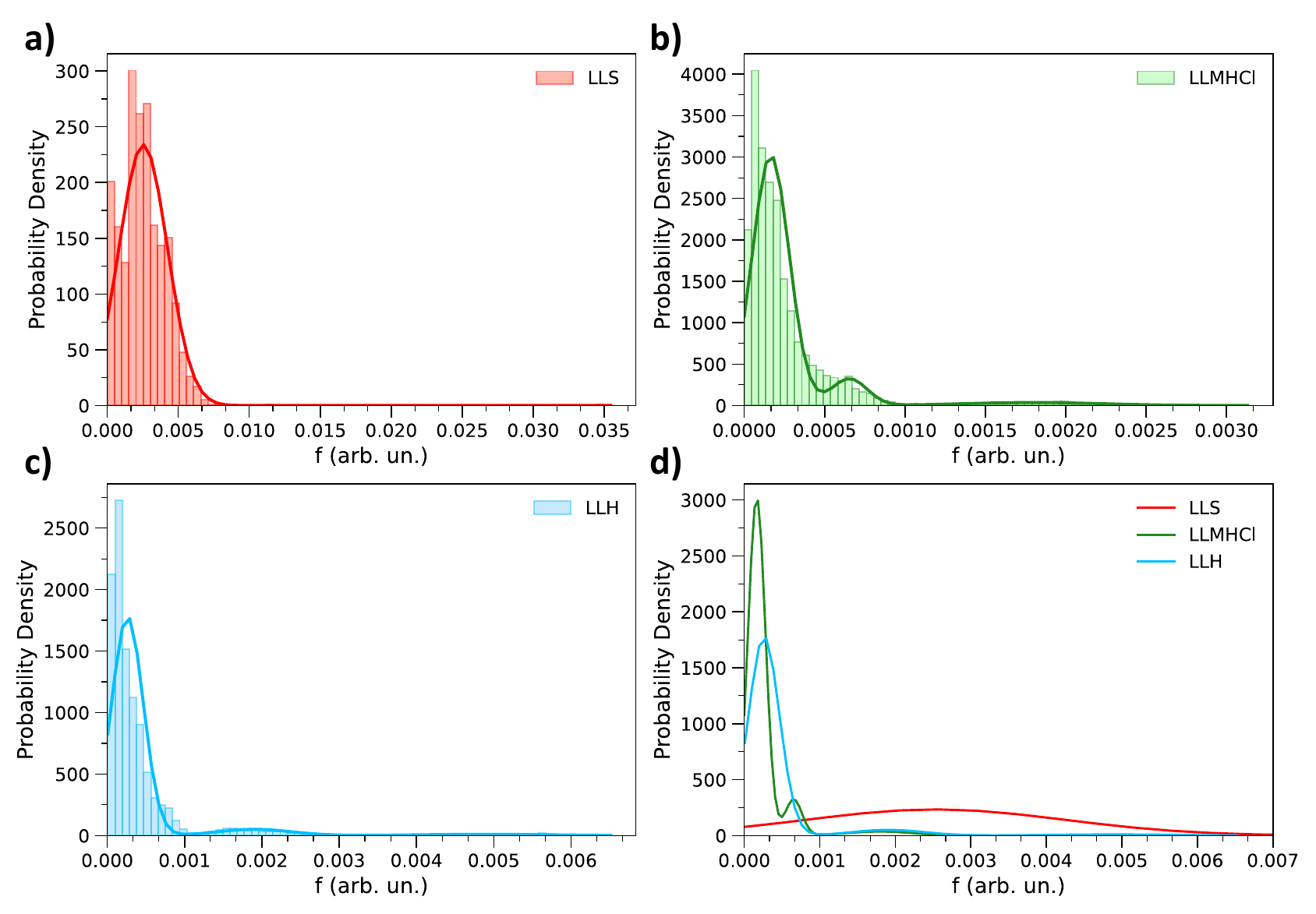}
    \caption{Normalized histograms of oscillator strength (f) distributions and corresponding Gaussian fits for the three \textup{\textsmaller{L}}-Lysine crystals: a) \textup{\textsmaller{L}}-Lysine sulfate (LLS), b) \textup{\textsmaller{L}}-Lysine monohydrochloride (LLMHCl), and c) \textup{\textsmaller{L}}-Lysine hemihydrate (LLH). Panel d) displays only the Gaussian fits from panels (a–c) for direct comparison of the probability density functions across different protonation states.}
    \label{fig:SI_11}
\end{figure}
\clearpage

\begin{figure}[H]
    \centering
    \includegraphics[width=\linewidth]{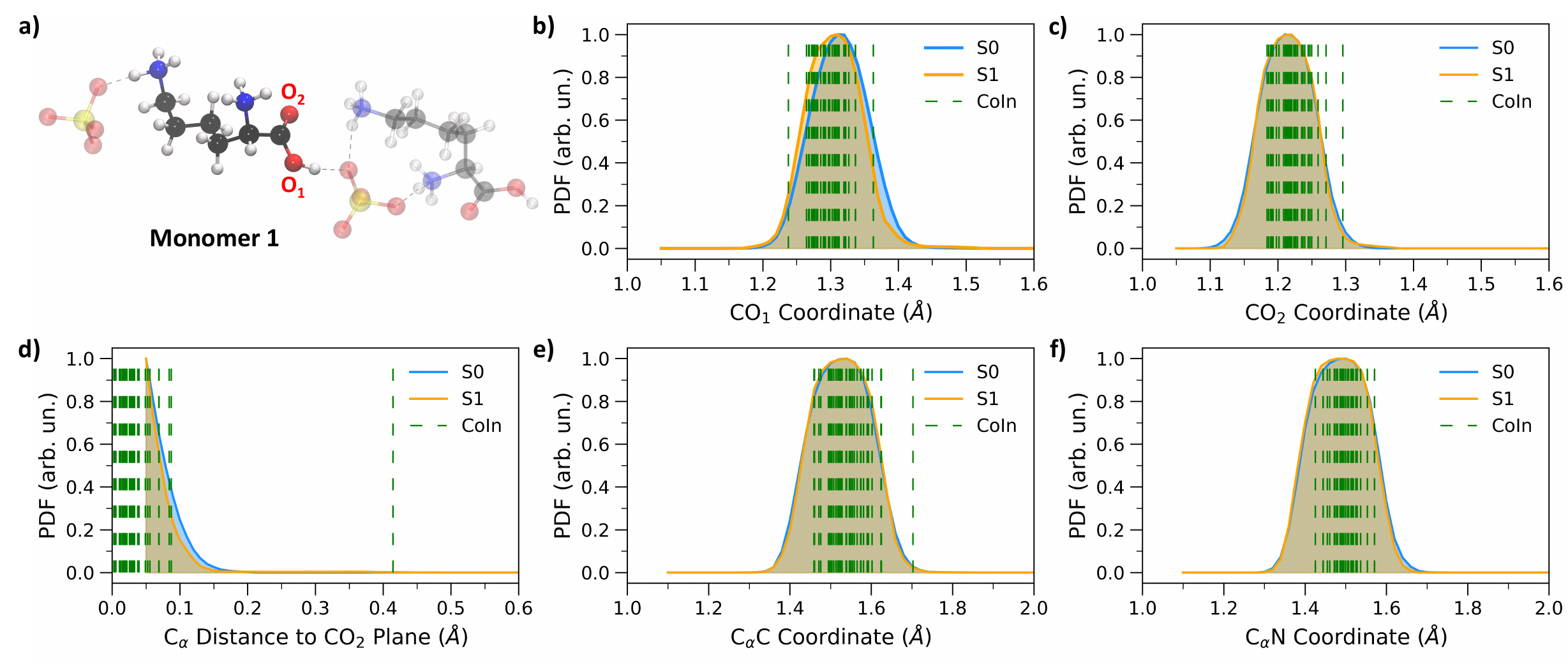}
    \caption{Probability distribution functions (PDFs) for selected vibrational coordinates of one of the two \textup{\textsmaller{L}}-Lysine sulfate (LLS) monomers within the quantum mechanics (QM) region (monomer 1, shown in panel a). Atoms of monomer 1 are depicted as glossy spheres, while the remaining atoms of the QM dimer are shown transparently. PDFs are reported for: (b, c) the two C=O stretching modes; (d) deplanarization of the C$_\alpha$–C–O–O fragment; (e) the C$_\alpha$C $\sigma$-bond stretch; and (f) the C$_\alpha$N $\sigma$-bond stretch. PDFs corresponding to the ground state S$_0$ are plotted in blue; those for the first excited state S$_1$ are plotted in orange. Dashed green vertical lines indicate the PT coordinate values at which the conical intersection (CoIn) is reached, and non-radiative decay occurs from S$_1\rightarrow$ S$_0$.}
    \label{fig:SI_12}
\end{figure}
\clearpage

\begin{figure}[H]
    \centering
    \includegraphics[width=\linewidth]{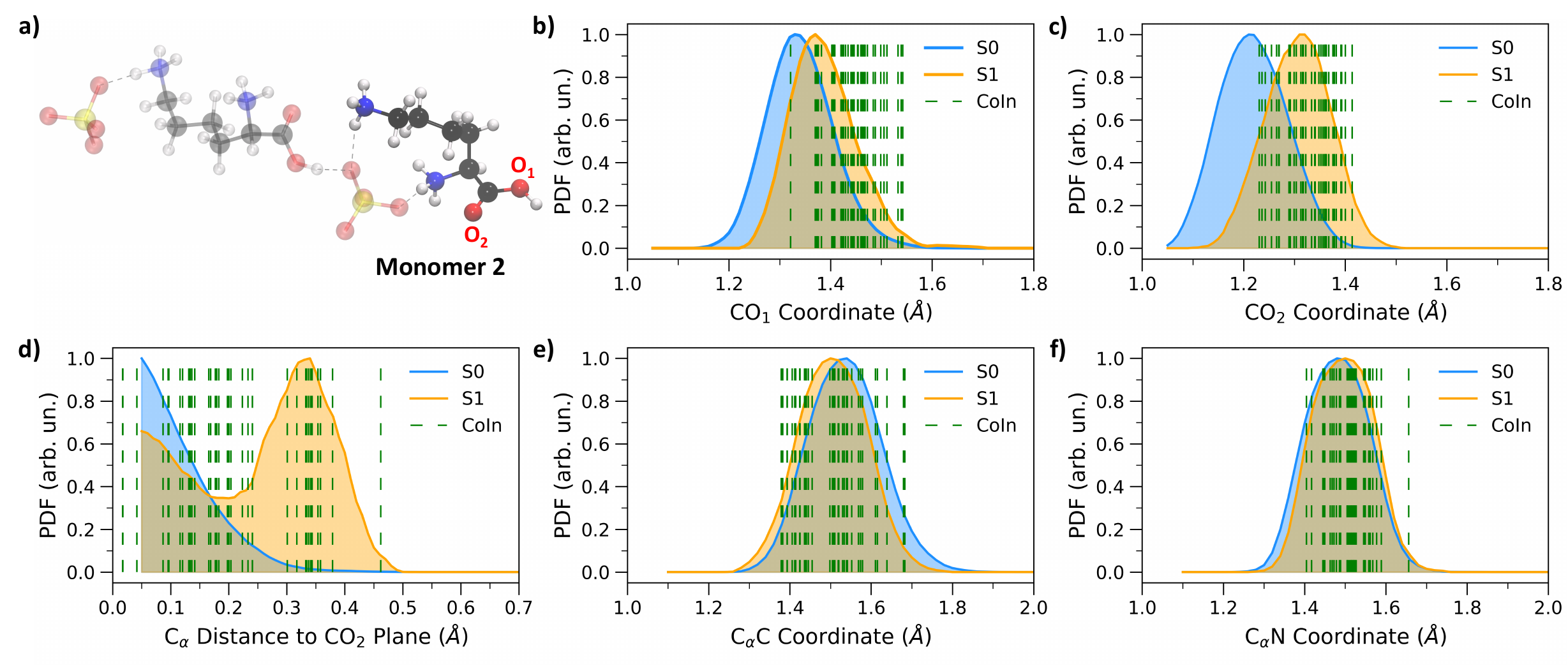}
    \caption{Probability distribution functions (PDFs) for selected vibrational coordinates of one of the two \textup{\textsmaller{L}}-Lysine sulfate (LLS) monomers within the quantum mechanics (QM) region (monomer 2, shown in panel a). Atoms of monomer 2 are depicted as glossy spheres, while the remaining atoms of the QM dimer are shown transparently. PDFs are reported for: (b, c) the two C=O stretching modes; (d) deplanarization of the C$_\alpha$–C–O–O fragment; (e) the C$_\alpha$C $\sigma$-bond stretch; and (f) the C$_\alpha$N $\sigma$-bond stretch. PDFs corresponding to the ground state S$_0$ are plotted in blue; those for the first excited state S$_1$ are plotted in orange. Dashed green vertical lines indicate the PT coordinate values at which the conical intersection (CoIn) is reached, and non-radiative decay occurs from S$_1\rightarrow$ S$_0$.}
    \label{fig:SI_13}
\end{figure}
\clearpage
\begin{figure}[H]
    \centering
    \includegraphics[width=\linewidth]{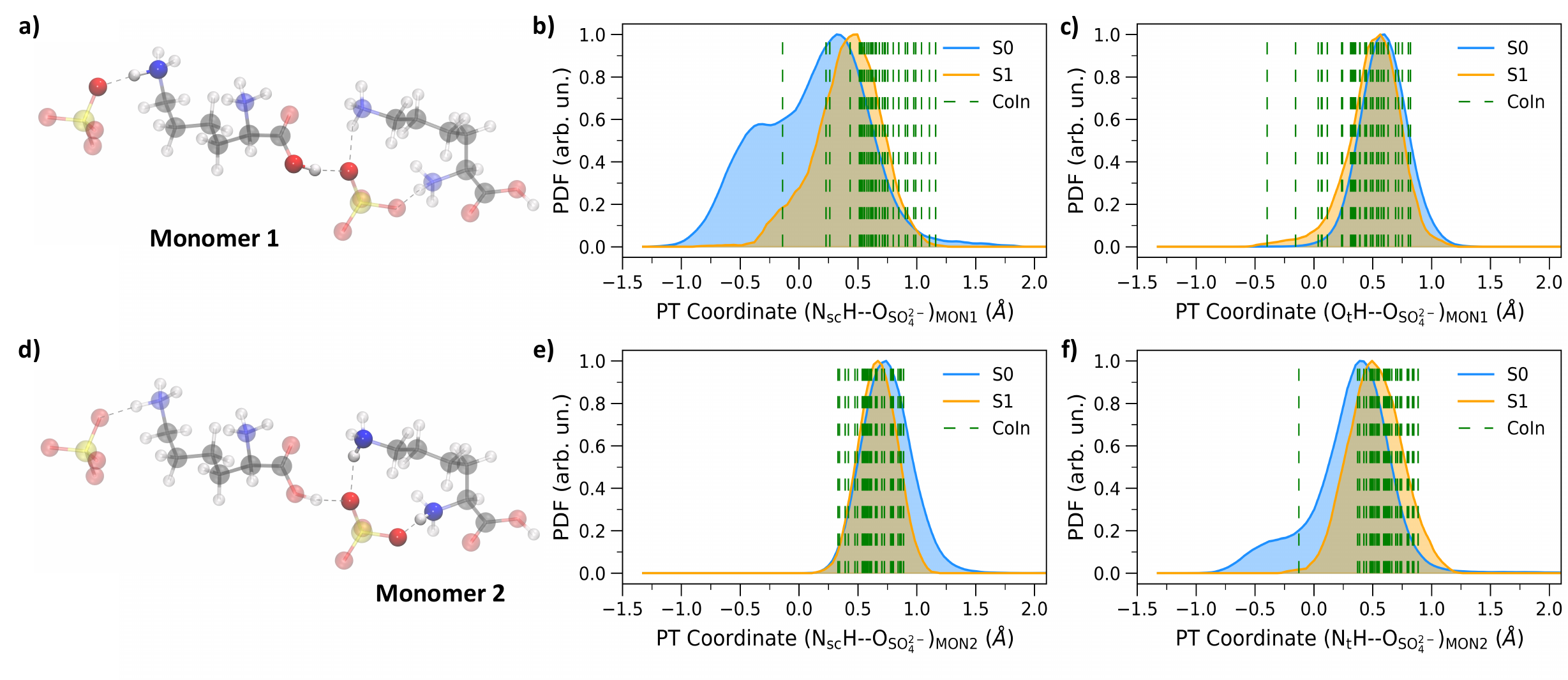}
    \caption{Probability distribution functions (PDFs) for the proton transfer (PT) coordinate, defined as $d_{\mathrm{X}\mathrm{H}} - d_{\mathrm{O}\mathrm{H}}$, where $\mathrm{X} = \mathrm{N}$ or $\mathrm{O}$ of an \textup{\textsmaller{L}}-Lysine sulfate (LLS) monomer treated at the quantum‑mechanical (QM) level, and $\mathrm{O}$ belongs to the SO$_4^{2-}$ counterion of the QM region. (a)–(c) refer to QM monomer 1 (MON1), while (d)–(f) refer to QM monomer 2 (MON2). Geometries of the dimers are shown in panels a) and d): atoms involved in the analyzed coordinates are shown as glossy spheres; all other atoms are rendered transparently. Panels b) and e) show the PT coordinate involving the side‑chain N atom of monomers 1 and 2, respectively. Panels c) and f) display the PT coordinate involving the terminal oxygen atom of monomer 1 and the terminal nitrogen atom of monomer 2. PDFs corresponding to the ground state S$_0$ are plotted in blue; those for the first excited state S$_1$ are plotted in orange. Dashed green vertical lines indicate the PT coordinate values at which the conical intersection (CoIn) is reached, and non-radiative decay occurs from S$_1\rightarrow$ S$_0$.
    }
    \label{fig:SI_14}
\end{figure}
\clearpage
\begin{figure}[H]
    \centering
    \includegraphics[width=\textwidth]{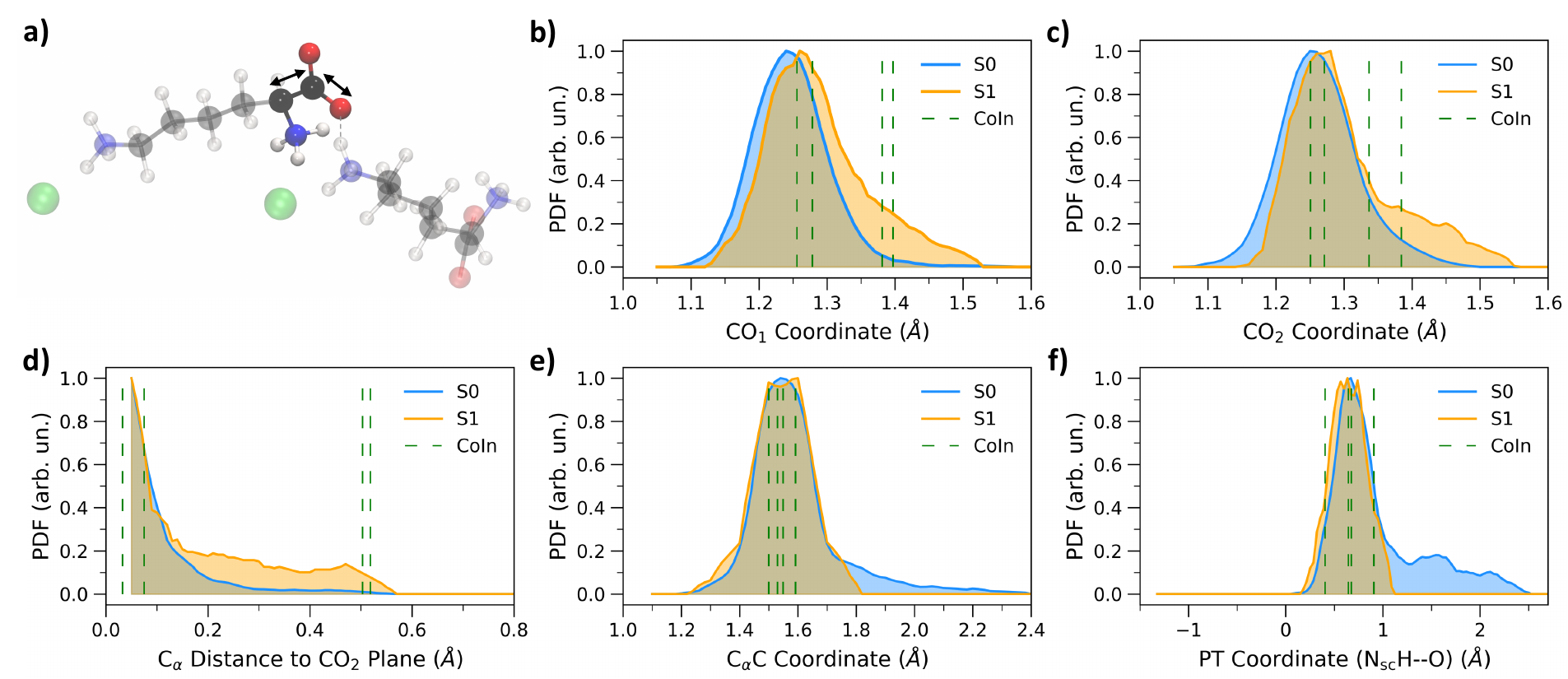}
    \caption{Probability distribution functions (PDFs) for the vibrational coordinates that govern non-radiative decay in four surface-hopping trajectories of \textup{\textsmaller{L}}-Lysine monohydrochloride (LLMHCl). Panel (a): geometry of the dimer treated at the quantum-mechanical (QM) level. Atoms involved in the analyzed coordinates are shown as glossy spheres; all other atoms are rendered transparently. PDFs are reported for: (b, c) the two C=O stretching modes; (d) deplanarization of the C$_\alpha$--C--O--O fragment; (e) the C$_\alpha$C $\sigma$-bond stretch; (f) the C$_\alpha$N $\sigma$-bond stretch. Similar results were observed for the corresponding coordinates in the other QM monomer. PDFs corresponding to the ground state S$_0$ are plotted in blue; those for the first excited state S$_1$ are plotted in orange. Dashed green vertical lines indicate the PT coordinate values at which the conical intersection (CoIn) is reached, and non-radiative decay occurs from S$_1\rightarrow$ S$_0$. 
    }
    \label{fig:SI_15}
\end{figure}
\clearpage

\begin{figure}[H]
    \centering
    \includegraphics[width=\textwidth]{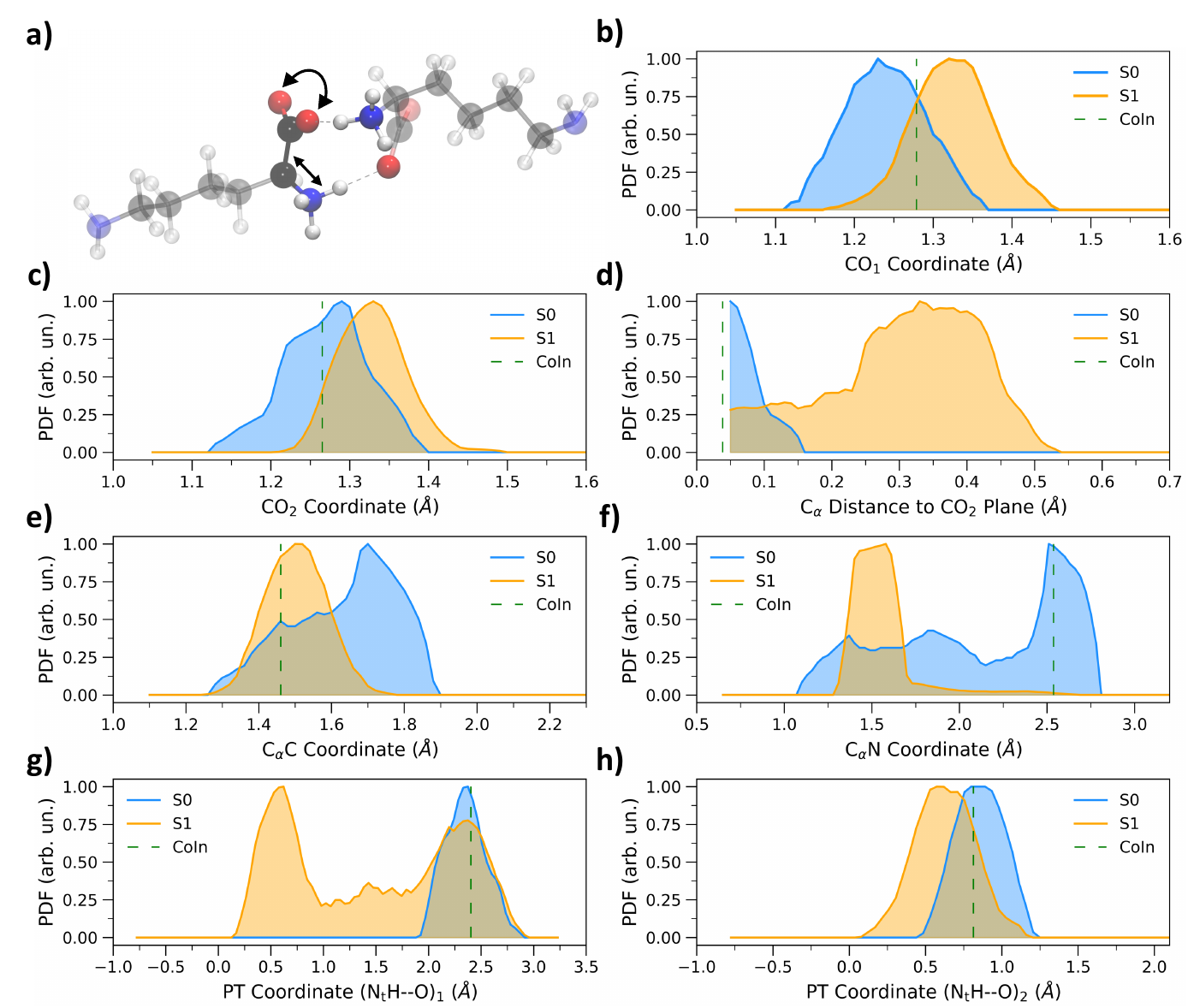}
    \caption{Probability distribution functions (PDFs) for the vibrational coordinates that govern non-radiative decay in a surface-hopping trajectory of \textup{\textsmaller{L}}-Lysine hemihydrate (LLH). Panel (a): geometry of the dimer treated at the quantum-mechanical level. Atoms involved in the analyzed coordinates are shown as glossy spheres; all other atoms are rendered transparently. PDFs are reported for: (b, c) the two C=O stretching modes; (d) deplanarization of the C$_\alpha$--C--O--O fragment; (e) the C$_\alpha$C $\sigma$-bond stretch; (f) the C$_\alpha$N $\sigma$-bond stretch; and (g, h) the two alternative proton transfer (PT) coordinates, defined as d$_{\text{N}_\text{t}\text{H}}-$d$_{\text{O}\text{H}}$. PDFs corresponding to the ground state S$_0$ are plotted in blue; those for the first excited state S$_1$ are plotted in orange. Dashed green vertical lines indicate the PT coordinate values at which the conical intersection (CoIn) is reached, and non-radiative decay occurs from S$_1\rightarrow$ S$_0$.
    }
    \label{fig:SI_16}
\end{figure}
\clearpage

\begin{figure}[H]
    \centering
    \includegraphics[width=\textwidth]{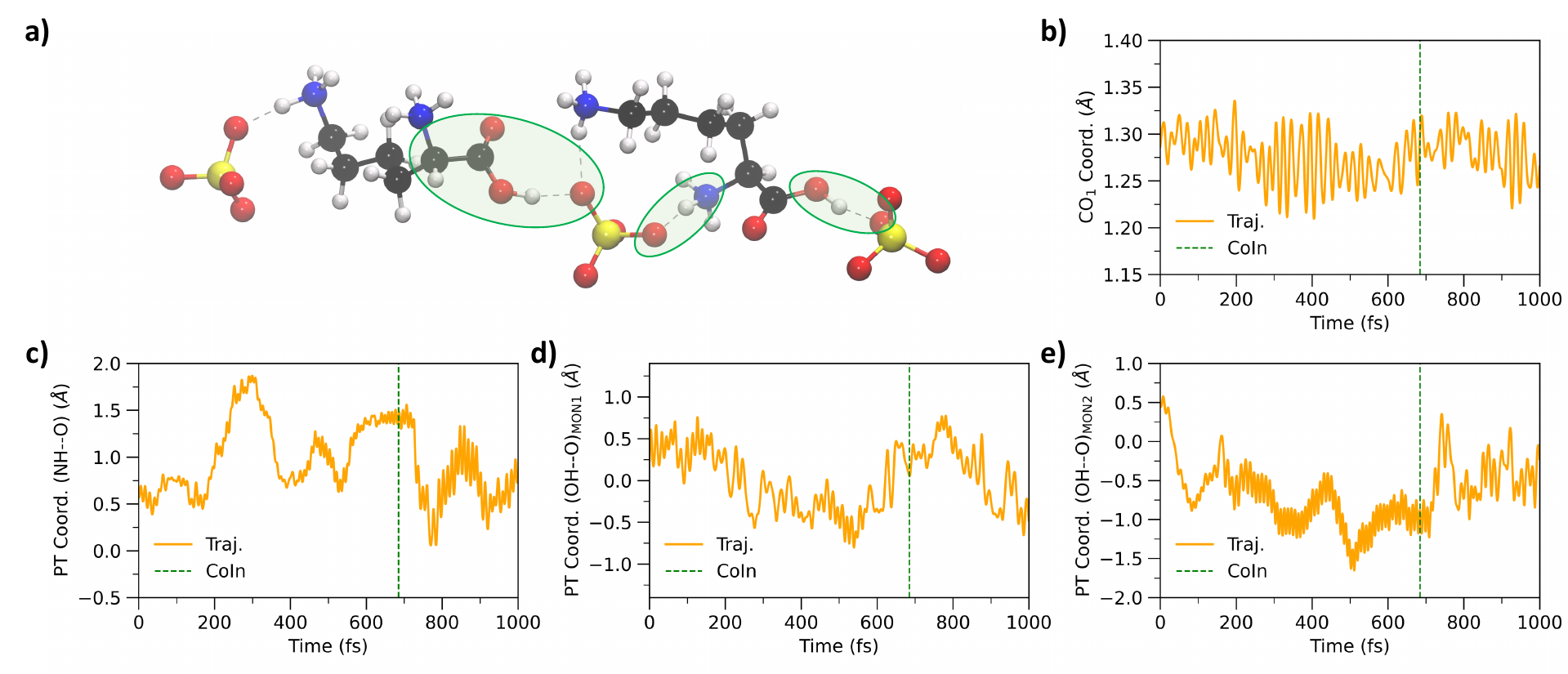}
    \caption{Time-series evolution of vibrational modes driving non-radiative decay in \textup{\textsmaller{L}}-Lysine sulfate (LLS) with an extended QM region. Panel (a) shows the QM subsystem used in this validation test, comprising a Lys dimer and three sulfate anions. This setup was employed to assess whether including an additional sulfate ion alters the decay mechanism observed in the standard QM/MM partitioning. Panels (b–e) show the time evolution (orange lines) of key vibrational coordinates for one representative trajectory (Traj.): (b) C=O coordinate (Coord.), and (c–e) proton transfer (PT) coordinates, defined as d${_{\text{X}\text{H}}}$ – d${_{\text{O}\text{H}}}$, where X = N (c) or O (d–e) from Lys, and O corresponds to the acceptor oxygen atom of SO$_4^{2-}$. Dashed green vertical lines indicate the mode values at which the conical intersection (CoIn) is reached, and non-radiative decay occurs from S$_1\rightarrow$ S$_0$.}
    \label{fig:SI_17}
\end{figure}
\end{suppinfo}

\clearpage
\bibliography{achemso-demo}

\end{document}